\documentclass[times,twocolumn,review]{elsarticle}

\usepackage{medima}
\usepackage{framed,multirow}

\usepackage{amssymb}
\usepackage{latexsym}
\usepackage{amsmath}
\usepackage{url}
\usepackage{xcolor}
\usepackage{adjustbox,subcaption}
\usepackage{hyperref}
\newcommand{\acc}{\textsf{acc}}
\newcommand{\conf}{\textsf{conf}}

\definecolor{newcolor}{rgb}{.8,.349,.1}

\journal{Medical Image Analysis}

\begin{document}

\verso{Ling  Huang \textit{et~al.}}

\begin{frontmatter}

\title{A review of uncertainty quantification in medical image analysis: probabilistic and non-probabilistic methods}%

\author[1]{Ling  \snm{Huang}}

\author[2]{Su \snm{Ruan}}
\cortext[cor1]{Corresponding author: Su Ruan,
  Email: su.ruan@univ-rouen.fr}
\author[1]{Yucheng  \snm{Xing}}
\author[1,3]{Mengling \snm{Feng}}

\address[1]{Saw Swee Hock School of Public Health, National University of Singapore, Singapore}
\address[2]{Quantif, LITIS, University of Rouen Normandy, France}
\address[3]{Institute of Data Science, National University of Singapore, Singapore}
\received{1 May 2013}
\finalform{10 May 2013}
\accepted{13 May 2013}
\availableonline{15 May 2013}
\communicated{S. Sarkar}

\begin{abstract}

The comprehensive integration of machine learning healthcare models within clinical practice remains suboptimal, notwithstanding the proliferation of high-performing solutions reported in the literature. A predominant factor hindering widespread adoption pertains to an insufficiency of evidence affirming the reliability of the aforementioned models. 
Recently, uncertainty quantification methods have been proposed as a potential solution to quantify the reliability of machine learning models and thus increase the interpretability and acceptability of the result. 
In this review, we offer a comprehensive overview of prevailing methods proposed to quantify uncertainty inherent in machine learning models developed for various medical image tasks.
Contrary to earlier reviews that exclusively focused on probabilistic methods, this review also explores non-probabilistic approaches, thereby furnishing a more holistic survey of research pertaining to uncertainty quantification for machine learning models.
%
Analysis of medical images with the summary and discussion on medical applications and the corresponding uncertainty evaluation protocols are presented, which focus on the specific challenges of uncertainty in medical image analysis. We also highlight some potential future research work at the end.
Generally, this review aims to allow researchers from both clinical and technical backgrounds to gain a quick and yet in-depth understanding of the research in uncertainty quantification for medical image analysis machine learning models.

\end{abstract}

\begin{keyword}
\KWD Uncertainty quantification
\sep Probabilistic methods 
\sep Non-probabilistic methods 
\sep Epistemic uncertainty 
\sep Aleatory uncertainty 
\sep Uncertainty evaluation  
\sep Medical image analysis 
\end{keyword}

\end{frontmatter}



\section{Introduction}
With the augmented investment of financial and human resources into artificial intelligence (AI), society has experienced notable transformations.
Healthcare is definitely one of the areas where we see great potential for AI to introduce revolutionizing improvements.
%
In particular, for medical image analysis (MIA), many deep neural network-based machine learning models with powerful learning and feature representation abilities have been developed. 
Despite the excellent performance of recent MIA methods, doubts about the reliability of their results still remain \citep{thagaard2020can, hullermeier2021aleatoric, czolbe2021segmentation}, which explains why their application to therapeutic decision-making for complex oncological cases is still limited. 
%
Learning, in the sense of generalizing beyond observed data so far, relies inherently on induction, i.e., replacing specific observations with general models of the data-generating process. Such models, however, are inherently speculative and lack definitive correctness; they remain hypothetical and, consequently, uncertain. The uncertainty extends to the predictions generated by these models as well.
In addition to the inductive inference uncertainty, other sources of uncertainty, such as incorrect model assumptions and noisy or imprecise data, exist, too.

\begin{figure}
    \centering
    \includegraphics[width=0.45\textwidth]{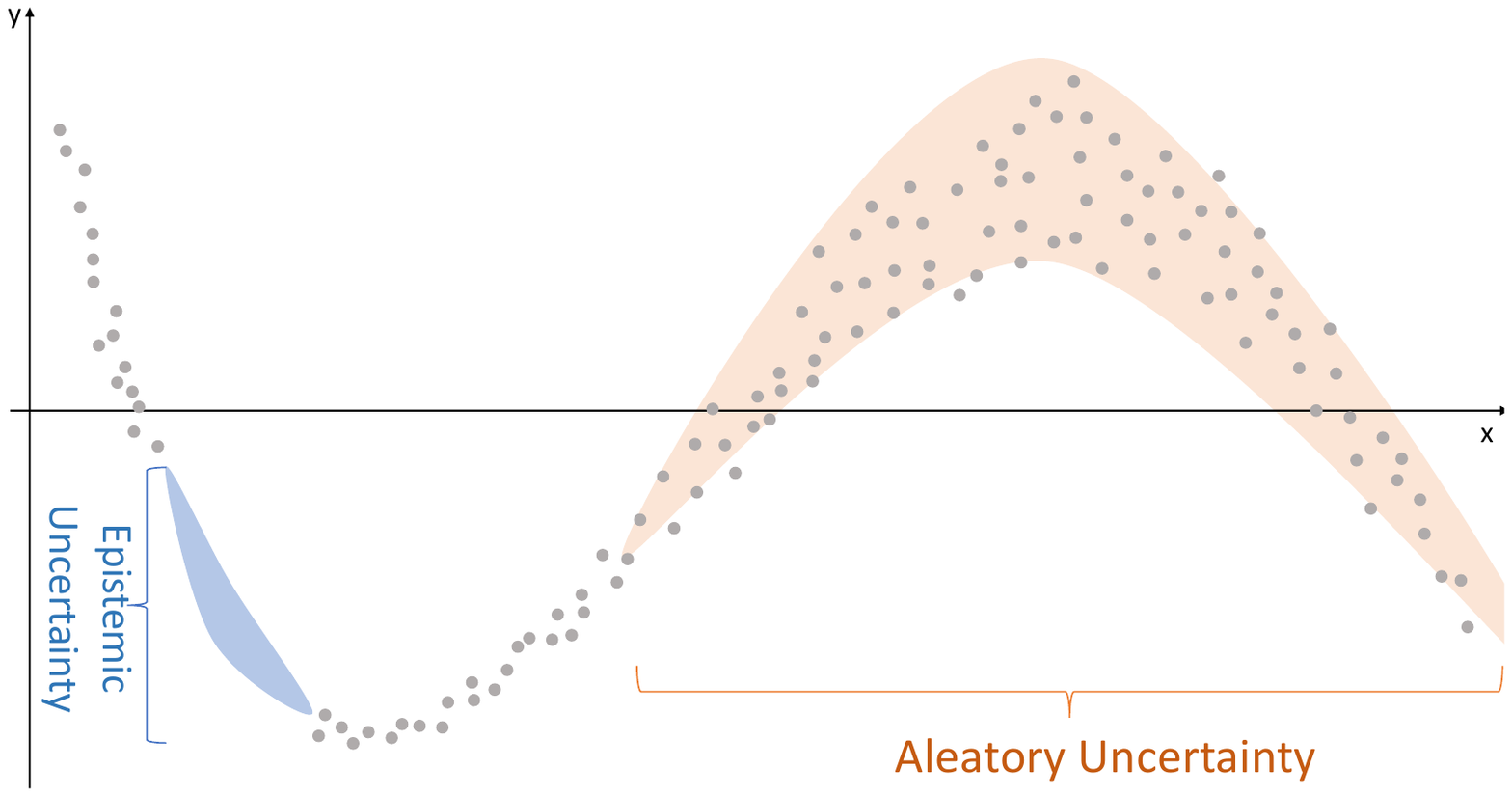}
    \caption{An example explanation of aleatory uncertainty inherently random and epistemic uncertainty  caused by a lack of knowledge about the best analysis model (reproduced based on \citep{yang2023explainable})}
    \label{fig: aleatoric and epistemic}
\end{figure}

In general, there are two sources of uncertainty: aleatory and epistemic uncertainty \citep{hora1996aleatory, der2009aleatory}. Aleatory uncertainty refers to the notion of randomness, i.e., the variability in an experimental outcome due to inherently random effects, which can not be reduced. In contrast, epistemic uncertainty refers to uncertainty caused by a lack of knowledge (ignorance) about the best analysis model, i.e., the ignorance of the learning algorithm or decision-maker. As opposed to uncertainty caused by randomness, uncertainty caused by ignorance can be reduced based on additional information or the design of a suitable learning algorithm. Figure \ref{fig: aleatoric and epistemic} provides an example explanation of aleatory and epistemic uncertainty.  

To fully harness the potential benefits of ML in MIA systems, a trustworthy representation of uncertainty is desirable and should be considered a key feature of developing state-of-the-art (SOTA) MIA methods. Traditionally, in fields like statistics and machine learning, probabilistic methods that rely on probability theory to represent, propagate, and analyze uncertainty have been perceived as the ultimate tool for uncertainty handling. They are commonly used with Bayesian inference to model uncertainty in various parameters or variables \citep{hinton1993keeping, mackay1992practical}. Moreover, the recent popularity of deep models has revived research on model uncertainty and has given rise to specific methods such as Monte Carlo dropout \citep{gal2016dropout, tran2018bayesian} and model ensembles \citep{lakshminarayanan2017simple, rupprecht2017learning}. However, probabilistic models make strong assumptions about the real distribution, which can potentially bring erroneous uncertainty estimation and fail to predict uncertainty correctly when the actual distribution is different. Moreover, probabilistic models are essentially based on capturing knowledge in terms of probability distribution and fail to distinguish between aleatory and epistemic uncertainty, limiting the exploitation of the results. 

Non-probabilistic methods, which handle uncertainty without relying on explicit probabilistic models, are usually used to characterize and analyze uncertainty when probabilistic information, such as precise probabilities or distributions, is unavailable or difficult to determine. Instead of building strong assumptions about the real distribution, these methods use alternative mathematical frameworks or representations such as intervals \citep{rao1997analysis}, fuzzy sets \citep{zadeh1965fuzzy}, Credal partition \citep{denoeux2004evclus}, or distance-based evidence reasoning mechanisms \citep{denoeux1995k} to quantify uncertainty. 

The recent study on uncertainty quantification significantly improved the performance of ML models and increased researchers' interest in analyzing those studies systematically. In 2016, Guney Gusel examined and explained fuzzy logic-based uncertainty methods in healthcare decision‑making \citep{gursel2016healthcare}. In 2018, Kabir et al. reviewed neural network-based uncertainty quantification methods with a particular focus on probabilistic forecasting and prediction intervals \citep{kabir2018neural}. In 2021, there are booming analyses about uncertainty. Alizadehsani et al. reviewed the research handling uncertainty in medical data using machine learning and probability theory techniques in the last 30 years \citep{alizadehsani2021handling}. Hüllermeier and Waegeman provided a comprehensive introduction to concepts and methods about aleatory and epistemic uncertainty in ML \citep{hullermeier2021aleatoric}. Abdar et al. reviewed uncertainty quantification in deep learning with discussions on techniques, applications and potential challenges with a particular focus on Bayesian statistics and ensemble learning \citep{abdar2021review}. Gillmann et al. studied uncertainty-aware visualization methods, showing readers which approaches can be combined to form uncertainty‐aware medical imaging pipelines \citep{gillmann2021uncertainty}. However, the above-mentioned review work can not provide a global overview of uncertainty quantification methods in MIA with recent ML methods, limiting the development of uncertainty analysis studies. 

\paragraph{Contributions}
Unlike previous uncertainty review papers that provide a general picture of uncertainty quantification in ML applications \citep{abdar2021review, hullermeier2021aleatoric}, or focus on discussing several specific uncertainty quantification methods \citep{alizadehsani2021handling}, this study reviews both probabilistic and non-probabilistic uncertainty methods in MIA under the ML framework in the last ten years, in which the later one is still ignored. It is worth mentioning that the primary purpose of this study is not to introduce the performance of various existing uncertainty quantification methods. Instead, we focus on outlining the most common uncertainty quantification and evaluation methods, the important application areas, as well as potential research work. We hope this review paper can provide guidance to researchers in the fields of machine learning and clinical practice and pave the way for future research in order to generate reliable and explainable decisions based on quantified uncertainty or improve the fairness of the overall healthcare systems by combining multiple source information with uncertainty. The main contributions of this study are as follows:
\begin{itemize}
\item To our best knowledge, this is the first comprehensive review paper regarding the study of both probabilistic and non-probabilistic uncertainty quantification methods in MIA tasks.
\item Existing uncertainty evaluation criteria applied for MIA are studied and discussed. 
\item The main categories of important clinical applications of uncertainty quantification methods are presented and discussed.
\item  The major advantages and limitations of existing uncertainty quantification research are pointed out, as well as the potential future work.
\end{itemize}

\paragraph{Organization} The rest of this paper is organized as follows: Section \ref{sec: criteria} explained the search criteria. Section \ref{sec: overview} presents the commonly used probabilistic and non-probabilistic uncertainty quantification methods in MIA. Section \ref{sec: methods} introduces the uncertainty evaluation criteria. Section \ref{sec: appli} summarizes MIA applications with the mentioned uncertainty quantification methods. Finally, Section \ref{sec: discuss} provides a discussion of the advantages and limitations of the literature, and Section \ref{sec: conclu} gives the overall conclusion of this review. 

\section{Search criteria}
\label{sec: criteria}
\begin{figure*}
    \centering
    \includegraphics[scale=0.5]{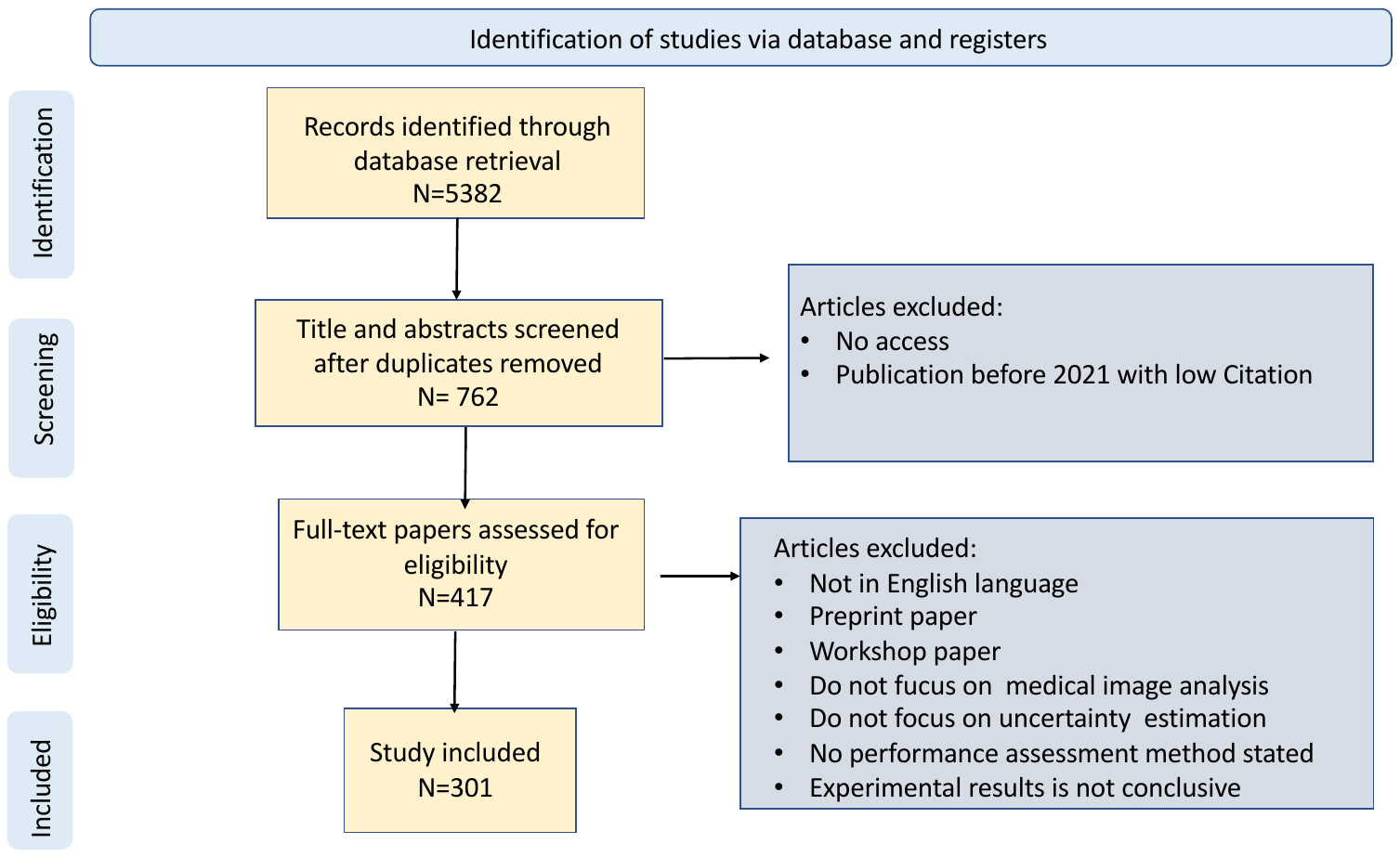}
    \caption{Illustration of selecting eligible publications for inclusion.}
    \label{fig: study-selection}
\end{figure*}
To perform this review, we performed a search on Web of Science for the papers published between 1 January 2013 and 15 July 2023. The search keywords used for this study are 'Uncertainty quantification' OR 'fuzzy systems' OR 'Monte Carlo simulation' OR 'rough classification' OR 'Dempster–Shafer theory' OR 'Imprecise probability' AND 'Medical image analysis.' We note that, from 2013 to 2023, more than 5,000 papers studying uncertainty in MIA tasks were published. To ensure the criticality of the study, we only include papers published in related journals and conferences and screen their title and abstracts. Then, about 700 papers with full access and good citation records were selected, and those lacking adequate connection with the topic of our review were removed from the list. Then, we read the full-text paper with the inclusion criteria illustrated in Figure \ref{fig: study-selection}. In the end, 301 papers are investigated in this review. 

\begin{figure}
\centering
\subfloat[Number of papers focused on uncertainty in ML\label{fig:class1}]{\includegraphics[width=0.5\textwidth]{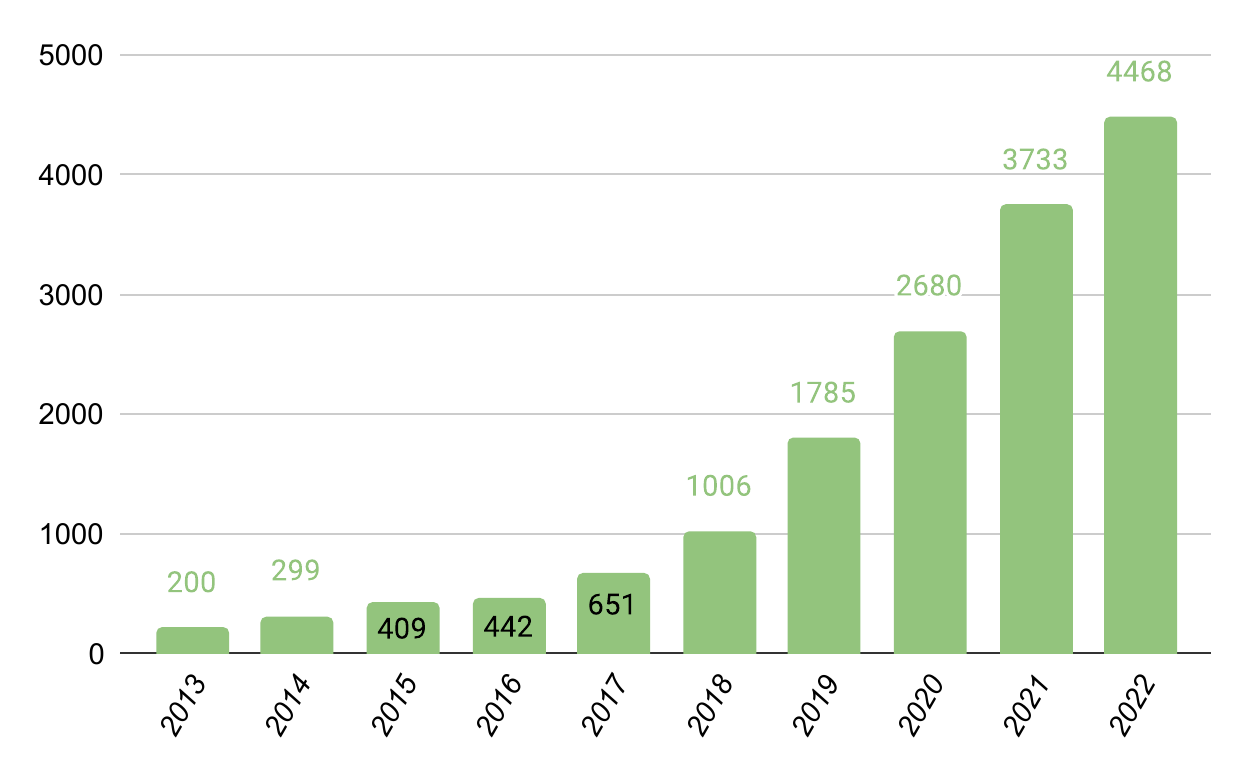}}\\
\subfloat[Number of papers focused on uncertainty in MIA\label{fig:class2}]{\includegraphics[width=0.5\textwidth]{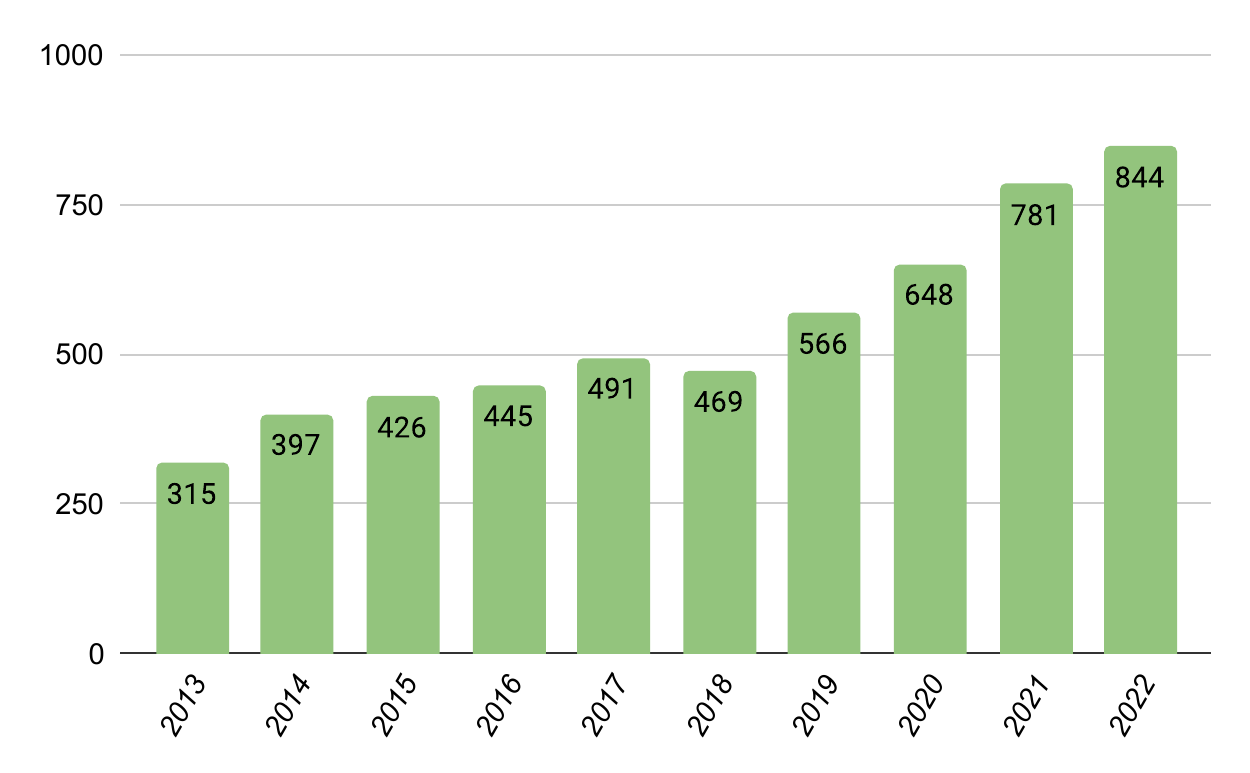}}
\caption{Number of papers focused on uncertainty in the last 10 years.}
\label{fig: year-number}
\end{figure}

Figure \ref{fig: year-number} shows the number of papers focused on uncertainty analysis in the last ten years, where we can see that handling uncertainty in machine learning has received increasing attention, especially when machine learning methods have been able to achieve promising accuracy performance with the popularization of deep learning after 2015. Researchers' interest in the study of uncertainty in MIA models remained at a steady state until 2018, i.e., around 400 published papers each year. There are two main reasons: 1) uncertainties in medical image reconstruction or registration tasks are easily observable and awarded; 2) the study of ML models for MIA is lagging behind ML research itself. Once the ML research for MIA has reached an accuracy saturation situation, people then turn to study uncertainty. Therefore, after 2018, increasing efforts have been involved in studying the MIA uncertainty.

\section{Methods to uncertainty quantification}
\label{sec: overview}
\begin{figure*}
    \centering
    \includegraphics[width=\textwidth]{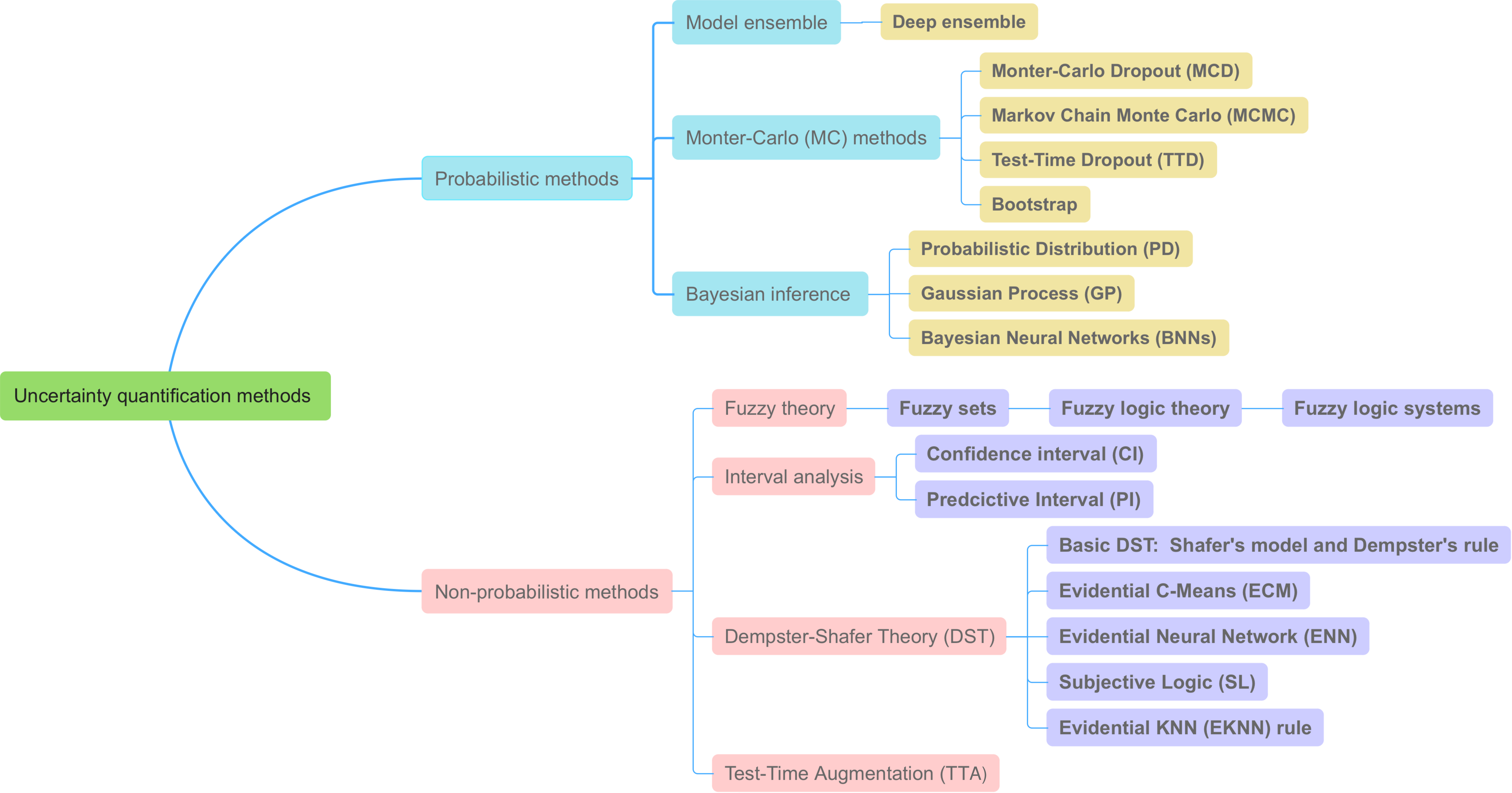 }
    \caption{Overview of uncertainty quantification methods: Probabilistic and Non-probabilistic methods}
    \label{fig: uncertianty-overview}
\end{figure*}

Different from existing literature reviews that focus on analyzing uncertainty methods in a specific field or with a specific methodology, in this paper, we provide a comprehensive overview of uncertainty analysis methods in medical images, including analysis from both probabilistic and non-probabilistic sides of the application to different medical image tasks. Figure \ref{fig: uncertianty-overview} shows an overview of uncertainty quantification methods. It should be noted that both probabilistic and non-probabilistic methods represent, propagate, and reason uncertainty in a systematic manner and the choice of method depends on the nature of uncertainty, available information, and the specific problem domain. 

\subsection{Probabilistic uncertainty quantification methods}
\label{subsec: pro}
Probabilistic uncertainty quantification methods leverage probability theory to represent uncertainty using probability distributions, allowing for calculating probabilities, quantiles, and other statistical measures. To capture uncertainty in ML model predictions, predictive entropy or variance are typically used to estimate distributions over the outputs. Predictive entropy \citep{shannon1948mathematical} measures the diversity or spread of a single probability distribution and quantifies how uncertain or ambiguous a model's prediction is by considering the distribution's entropy, which reflects the amount of information or randomness in the distribution, i.e., combining both aleatory and epistemic uncertainties into a single measure that reflects the overall uncertainty in a probabilistic prediction. Higher predictive entropy indicates greater uncertainty and ambiguity in the model's prediction. Predictive variance \citep{fisher1919xv} is a measure of uncertainty related to the inherent randomness or noise in the data, i.e., aleatory uncertainty. Higher predictive variance suggests that individual predictions are more scattered around the mean prediction, indicating that the model's predictions are more sensitive to changes in the input. According to our literature review, there are three main probabilistic methods: \emph{Bayesian Inference, Monte-Carlo method, and Model ensemble}, which can generate predictive entropy or variance.

\emph{Bayesian inference} is a statistical approach that inherently involves probabilistic modeling and updates prior beliefs or knowledge to estimate uncertain quantities using observed data. It combines prior distributions with likelihood functions to obtain posterior distributions that reflect updated beliefs. Among Bayesian inference methods, Probabilistic Distribution (PD) \citep{wallman2014computational}, Gaussian Process (GP) \citep{wachinger2014gaussian} and Bayesian Neural Networks (BNNs) \citep{blundell2015weight} are three main popular methods used for uncertainty quantification. Readers can refer to Supplementary Material A for detailed analysis. 

\emph{Monte Carlo (MC)} methods \citep{kroese2014monte} involves generating random samples from probability distributions. As the number of samples increases, the simulated outcomes converge toward the true distribution of possible outcomes, allowing us to obtain accurate estimates of uncertainty. Aleatory uncertainty is then captured through the variability introduced by dropout during forward passes, reflecting data randomness. Epistemic uncertainty is captured through the diversity of predictions across passes, indicating the model's uncertainty about its own parameters and structure. Among the MC methods, Monte Carlo sampling, Monte Carlo dropout (MCD) \citep{gal2016dropout}, Markov Chain Monte Carlo (MCMC) \citep{gilks1995markov, brooks1998markov}, and Bootstrap are the most common algorithms. Since Test-Time Dropout (TTD) \citep{srivastava2014dropout} also involves repeated sampling from the data, here we classify it into the MC methods as well (details can be found in Supplementary Material A). It should be noted that while the MC methods are not inherently a Bayesian inference method, it is often employed in Bayesian inference to estimate posterior distributions and perform various Bayesian analyses. 

\emph{Model ensemble} \citep{dietterich2000ensemble, zhou2012ensemble} typically focuses on capturing different sources of variability or uncertainty in model assumptions rather than explicitly quantifying uncertainty using probabilistic measures. Each model in the ensemble framework may be trained using different initializations, subsets of the training data, or variations in the model architecture. Standard ensemble methods, such as bagging, boosting, or random forests, generate an ensemble of models that collectively represent uncertainty with the variability among the predictions. Recently, deep ensemble \citep{lakshminarayanan2017simple, ganaie2022ensemble} models have become a popular uncertainty quantification method integrated with deep neural networks. Details about deep ensemble models can be found in Supplementary Material A.

To sum up, probabilistic uncertainty quantification methods provide a rigorous and quantitative approach to characterizing and analyzing uncertainty. 

\subsection{Non-probabilistic uncertainty quantification methods}
\label{subsec: nonpro}
Non-probabilistic uncertainty methods, free from the strong assumption of the prior distribution of the data, are more flexible and applicable for most applications, especially when precise probabilistic information is not available. Methods such as \emph{interval analysis} \citep{rao1997analysis}, \emph{fuzzy sets and fuzzy logic theory} \citep{yager2012introduction}, \emph{Dempster-Shafer theory} \citep{dempster1967upper, shafer1976mathematical}, and \emph{Test-Time Augmentation} do not directly involve probability distributions to represent uncertainty. Instead, they introduce conceptions such as predictive intervals \citep{eaton2018towards}, fuzzy membership functions \citep{wang2023novel} or plausible sets \citep{adiga2022leveraging}, Credal partition \citep{denoeux2004evclus}, evidence-based reasoning mechanisms \citep{huang2021evidential} or test-time data augmentation to model uncertainty. 

\emph{Dempster-Shafer theory (DST)} \citep{dempster1967upper, shafer1976mathematical}, also known as Belief function theory or Evidence theory, was first originated by Dempster \citep{dempster1967upper} in the context of statistic inference in 1968 and was formalized by Shafer \citep{shafer1976mathematical} as a theory of evidence in 1976. It is a theoretical framework for modeling, reasoning with, and combining imperfect (imprecise, uncertain, and partial) information. With DST, we can quantify uncertainty in a single forward pass and further explore the possibility of improving the model reliability based on the quantified uncertainty. Based on DST, there are some commonly used uncertainty quantification methods, i.e., Evidential KNN ({EKNN}) rule \cite{denoeux1995k}, Evidential C-Means (ECM) \citep{masson2008ecm}, Evidential Neural Network (ENN) \citep{denoeux2000neural}, and Subjective Logic (SL) \citep{josang2006trust, josang2016subjective}, readers can refer to Supplementary Material or paper \citep{HUANG2023737} for more information. 

\emph{Fuzzy sets} \citep{zadeh1965fuzzy} define the linguistic terms and their membership functions; fuzzy rules capture the relationships between inputs and outputs using if-then statements; and the fuzzy inference mechanism combines the rules and performs fuzzy reasoning to compute the system's output \citep{dubois1980fuzzy}. Fuzzy logic \citep{hajek2013metamathematic} employs fuzzy sets to capture the degree of membership of elements to a particular linguistic variable such as "high likelihood," "medium uncertainty," or "low confidence." \emph{Fuzzy logic systems} \citep{mendel1995fuzzy, yager2012introduction} are the implementations of fuzzy logic principles to solve specific problems by utilizing fuzzy sets, fuzzy rules, and fuzzy inference mechanisms. 

\emph{Interval analysis} \citep{rao1997analysis} represents uncertainty by bounding the possible range of values for variables or parameters using intervals, thus offering a systematic and robust approach to uncertainty quantification, especially in cases where rigorous bounds on uncertainty are essential for decision-making or risk assessment. In interval analysis, uncertainty is characterized by assigning intervals to model parameters, inputs, or outputs rather than specifying precise probabilities. These intervals represent ranges of possible values rather than probabilities of occurrence. Thus, it can be defined based on available information, expert opinion, or experimental data. Confidence intervals (CI) \citep{hosmer1992confidence, smithson2003confidence} and prediction intervals (PI) \citep{hwang1997prediction} are two common interval algorithms used to quantify the uncertainty associated with a given estimate. However, it can also lead to wide intervals if the input uncertainties are too large or if the model's behavior is nonlinear and complex.

\emph{Test-Time Augmentation (TTA)} \citep{ayhan2018test, wang2019aleatoric} is a technique used in machine learning to improve model performance and enhance the robustness of predictions. At test time, multiple variants of the input image are generated using data augmentation such as spatial transformations (e.g., flipping, rotation), intensity augmentations (e.g., contrast modification, noise injection, or artifacts), etc. Using TTA, the model generates a set of predictions for the same initial input image. From this distribution of predictions, uncertainty metrics can be extracted, such as the median or variance.


It should also be noted that there are some researches that hybrid more than one uncertainty quantification method, e.g., integrating MCD in deep ensemble models or integrating fuzzy set with DST, for uncertainty analysis and yield more promising performance. The detailed applications of the above-mentioned methods will be introduced in Section \ref{sec: appli}.

\section{Methods to uncertainty evaluation}
\label{sec: methods}
The previous section presented the main uncertainty estimation approaches applied to MIA tasks. In this section, we introduce the protocols implemented in these papers to evaluate the performance of the uncertainty estimation approaches. 

Direct uncertainty evaluation methods such as mean square error validate the correctness of uncertainty quantification techniques with given uncertainty ground truths. While in real-world medical scenarios, ground-truth uncertainty is unavailable or difficult to obtain. 

Indirect uncertainty evaluation methods, e.g., calibration metrics, coverage metrics, scoring rules, and prediction entropy, on the other hand, focus on a qualitative assessment of the computed uncertainty estimates by evaluating how well their predicted uncertainties correlate with the actual outcomes or data variability when uncertainty ground truth is unavailable. Misclassification or Out-of-Distribution detection are downstream applications of uncertainty in an automated pipeline of prediction, thus also used quite often in assessing the quality of the uncertainty quantification model. According to the literature review, we grouped five common uncertainty evaluation protocols (see Table \ref{tab: evaluation}). 
\begin{table*}
  \centering
  \caption{Evaluation criteria}
  \scalebox{0.9}{
  \begin{tabular}{lllllll}
 \hline
Evaluation criteria& Papers \\
\hline
\multirow{5}{*}{Coverage metrics}& \cite{judge2022crisp, mehta2023evaluating, nair2020exploring, valen2022quantifying, camarasa2021quantitative}\\
& \cite{qian2020cq, mehta2021propagating, eaton2018towards, bian2020uncertainty, kwon2020uncertainty}\\
&\cite{yang2021exploring, wickstrom2020uncertainty, wallman2014computational, ebadi2023cbct, kushibar2022layer}\\
&\cite{herzog2020integrating, le2016quantifying, gour2022uncertainty, corrado2020quantifying, arega2023automatic}\\
&\cite{jafari2021u, balagopal2021deep, awate2019estimating, risholm2013bayesian, kabir2022aleatory}\\
\hline
\multirow{4}{*}{Predictive Entropy}&  \cite{hamedani2023breast, jungo2018effect, mehta2023evaluating, mehta2021propagating, wang2021medical}\\
 &\cite{del2023labeling, gour2022uncertainty, ghoshal2020estimating, narnhofer2021bayesian}\\ 
 &\cite{ebadi2023cbct, kushibar2022layer, camarasa2021quantitative, rajaraman2022uncertainty, qian2020cq}\\
&\cite{herzog2020integrating, dai2013entropy, arega2023automatic}\\
\hline
\multirow{6}{*}{Calibration metrics}&\cite{hamedani2023breast, jungo2019assessing, sambyal2022towards,  laves2021recalibration}\\
& \cite{judge2022crisp, carneiro2020deep, liao2019modelling, ayhan2020expert, islam2021spatially}\\
& \cite{dawood2023uncertainty, thagaard2020can, pandey2022can, javadi2022towards}\\
& \cite{ghoshal2022calibrated, laves2021recalibration, ghoshal2021cost, carneiro2020deep}\\
& \cite{ dawood2023uncertainty, buddenkotte2023calibrating, mehrtash2020confidence, rousseau2021post}\\
& \cite{li2022region, jena2019bayesian, arega2023automatic}\\
\hline
\multirow{3}{*}{Misclassification detection \& OoD} & \cite{hamedani2023breast, jungo2019assessing, devries2018leveraging} \\
& \cite{ghoshal2019estimating, iwamoto2021improving, belharbi2021deep, asgharnezhad2022objective} \\
&\cite{linmans2023predictive, fuchs2021practical, thagaard2020can}\\
\hline
\multirow{2}{*}{Scoring functions}  &\cite{sambyal2022towards, mehrtash2020confidence, arega2023automatic, thagaard2020can}\\
 & \cite{mehrtash2020confidence, tanno2017bayesian, thagaard2020can, lemay2022label}\\
\hline

\hline

\end{tabular}
}
\label{tab: evaluation}
\end{table*}

\subsection{Coverage metrics}
Coverage metrics measure the proportion of cases where predicted uncertainty intervals (e.g., confidence intervals) contain the true value or the average width of prediction intervals. It can be estimated by sample variance or coverage probability. Sample variance computes the output variance across all samples collected using Bayesian inference, MC methods, or model ensembles with the definition:
\begin{equation}
    variance=\sqrt{\frac{\sum_{n=1}^{N}(y_n-\bar{y} )^2 }{N-1} },
\end{equation}
where $y_n$ is the value of the observation corresponding to pixel/voxel $n$, $\bar{y}$ is the mean value of all observations, and $N$ is the number of observations. Coverage probability measures the proportion of true outcomes within the predicted uncertainty intervals \citep{dodge2003oxford}. The construction of the confidence interval ensures that the probability of finding the true vector $\theta$ in the sample dependent interval $[T_u, T_v]$ is (at least) $\gamma$:
\begin{equation}
    P (T_u<\theta<T_v)=\gamma
\end{equation}

\subsection{Predictive entropy}
The predictive entropy measures the informativeness of the model’s predictive density function for each model output $y_i$ with the definition
\begin{equation}
    Entropy=-\sum_{i=1}^{C} p(i)\log p(i),
\end{equation}
where $p(i)$ denotes the probability density function (PDF) \citep{parzen1962estimation} or probability mass function (PMF) \citep{stewart2009probability} of the predicted variable $i$, and $C$ is the set of possible values for the predicted variable.

\subsection{Calibration metrics}
Calibration metrics measure the agreement between the predicted uncertainty and the observed frequency of correct predictions. A well-calibrated uncertainty estimation method should provide uncertainty estimates that align with the true error level or uncertainty in the predictions. Calibration can be assessed using calibration plots, reliability diagrams, or calibration metrics such as Calibration Error (CE), Maximal Calibration Error (MCE), and Expected Calibration Error (ECE). Here, we introduce ECE as an example. It measures the correspondence between predicted probabilities and ground truth \citep{guo2017calibration}. The output normalized plausibility of the model is first binned into equally spaced bin $E_h$, $h \in [1, H]$. The accuracy of bin $E_h$ is defined as
 \begin{equation}
     \acc(E_h)=\frac{1}{\mid E_h\mid}\sum_{n \in E_h}^{} \boldsymbol{1} (S_{n}=G_{n}),
 \end{equation}
 where $S_n$ and $G_n$ are, respectively, the predicted and true class labels for pixel/voxel $n$. The average confidence of bin $E_h$ is defined as
 \begin{equation}
\conf(E_h)=\frac{1}{\mid E_h \mid}\sum_{n \in E_h} ^{} P_{n},
 \end{equation}
 where $ P_{n}$ is the predicted probability for pixel/voxel $n$. The ECE is the weighted average of the difference in accuracy and confidence of the bins:
\begin{equation}
ECE= \sum_{h=1}^{H} \frac{\mid E_h \mid }{N}\mid \acc(E_h)-\conf(E_h)\mid,
\label{eq:ece}
\end{equation}
where $N$ is the total number of pixels/voxels in all bins here, $\mid E_h\mid$ is the number of elements in bin $E_h$. A model is perfectly calibrated when $\acc(E_h)=\conf(E_h)$ for all $h\in \{1,...,H\}$. 

\subsection{Scoring functions}
Brier score \citep{brier1950verification} and Negative log-likelihood (NLL) are two commonly used scoring functions for evaluating the performance of uncertainty estimation methods. Brier score \citep{brier1950verification} measures the mean squared difference between predicted probabilities and actual outcomes with:  
\begin{equation}
    BS=\frac{1}{N} \sum_{n=1}^{N} (P_{n}-G_{n})^2,
    \label{eq: bs}
\end{equation}
where $G_n$ is the ground truth of pixel/voxel $n$ and $P_{n}$ is the predicted probability of pixel/voxel $n$, $N$ is the number of pixels/voxels here. The lower the Brier score, the better the calibration and accuracy of the uncertainty estimates. NLL is usually used for evaluating probabilistic models and assessing their calibration and accuracy in capturing the uncertainty in predictions. It measures the average log probability assigned by a model to the observed outcomes by:
\begin{equation}
    NLL=-\sum_{n=1}^{N}G_n log(P_n) +(1-G_n)log(1-P_{n}),
    \label{eq: nll}
\end{equation}
where $G_n$ is the ground truth of pixel/voxel$n$ and $P_{n}$ is the predicted probability of pixel/voxel $n$, $N$ is the number of pixels/voxels here. A lower NLL value indicates better calibration and accuracy of the uncertainty estimates.

\subsection{Misclassification \& Out-of-Distribution  detection protocol}
A direct downstream application of uncertainty in an automated pipeline is the detection of samples for which the prediction is likely to be incorrect or Out-of-Distribution (OoD). This is crucial to prevent silent errors that could have a dramatic impact, especially in real-world medical image applications. In that sense, the uncertainty estimates can be turned into a binary classifier that aims at distinguishing between correct and incorrect predictions. As in the binary classification setting, an uncertainty threshold is applied to distinguish between positive (i.e., certain) and negative (i.e., uncertain) samples. The result of this classification is then compared to the true label of each sample, namely correct or incorrect. In that context, a confusion matrix \citep{stehman1997selecting} can be constructed from the uncertainty point of view by distinguishing four possible cases with the following counts: 
\begin{itemize}
    \item True Positive (TP): The prediction is uncertain, and the expected label and the prediction differ,
    \item False Negative (FP): The prediction is certain, but the expected label and the prediction differ, 
    \item True Negative (TN): The prediction is certain, and the expected label and the prediction are identical, 
    \item False Negative (FN): The prediction is uncertain, but the prediction and the expected label are identical.
\end{itemize}
\subsection{Discussion}
It's important to note that different evaluation metrics and approaches may be suitable depending on the context and specific application. To evaluate the performance of uncertainty estimation methods, it is necessary to employ additional quantitative evaluation measures, such as calibration, coverage probability, mean squared error, discrimination metrics, or task-specific performance metrics. These metrics assess the uncertainty estimates' accuracy, calibration, and discriminative ability. Apart from the numerical uncertainty evaluation, a visual inspection of uncertainty \citep{sedai2018joint, gillmann2020uncertainty} is also a valuable tool that is usually performed to verify whether they correspond to cases that a human would consider uncertain and usually be used for exploring, interpreting, and communicating uncertainty. 

The most controversial point of the current research method is the lack of uncertainty ground truth. With ground truth uncertainty, measuring the exact agreement between estimated and actual uncertainties becomes possible, while the lack of ground truth makes it challenging to assess the accuracy and calibration of uncertainty estimates quantitatively, reducing the development of uncertainty evaluation methods. 

\section{Applications of uncertainty quantification in MIA}
\label{sec: appli}
\begin{figure*}
    \centering
    \includegraphics[scale=0.5]{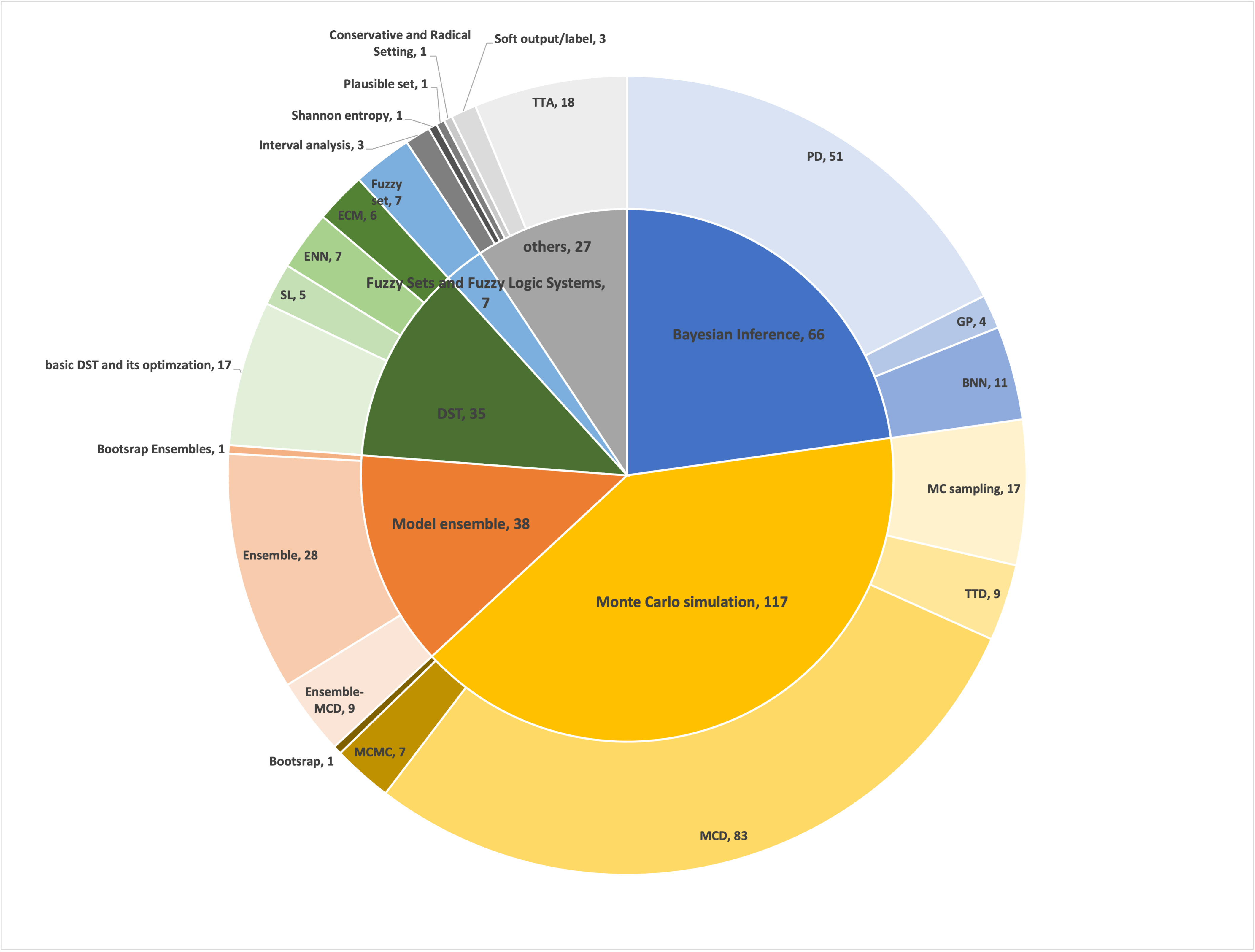}
    \caption{Statistics of uncertainty methods used in medical image analysis}
    \label{fig: method-classification}
\end{figure*}

The application of uncertainty quantification can help increase the accuracy of different MIA tasks. In MIA tasks, the uncertainty can be decomposed into three levels \citep{lakshminarayanan2017simple}: pixel/voxel-level, instance-level and subject-level. \emph{Pixel/voxel-level} uncertainty quantification is useful for interaction with physicians by providing additional guidance for correcting reconstruction/registration/segmentation results. \emph{Instance-level} uncertainty is the uncertainty aggregated by a set of pixel/voxel-level uncertainty. Its quantification can be used to reduce the false discovery rate for detection/prediction/classification tasks. \emph{Subject-level} uncertainty offers information on whether or not the model is about a subject. Therefore, quantifying uncertainty in MIA tasks is a critical step in advancing the field of medical imaging, allowing for better decision-making, fostering continual improvement of algorithms and risk assessment, and promoting transparency and trust between experts and patients, as well as ensuring safe and effective healthcare practices. Figure \ref{fig: method-classification} shows the overall statistics of probabilistic and non-probabilistic uncertainty methods used in MIA tasks. 

\begin{figure}
    \centering
    \includegraphics[width=0.5\textwidth]{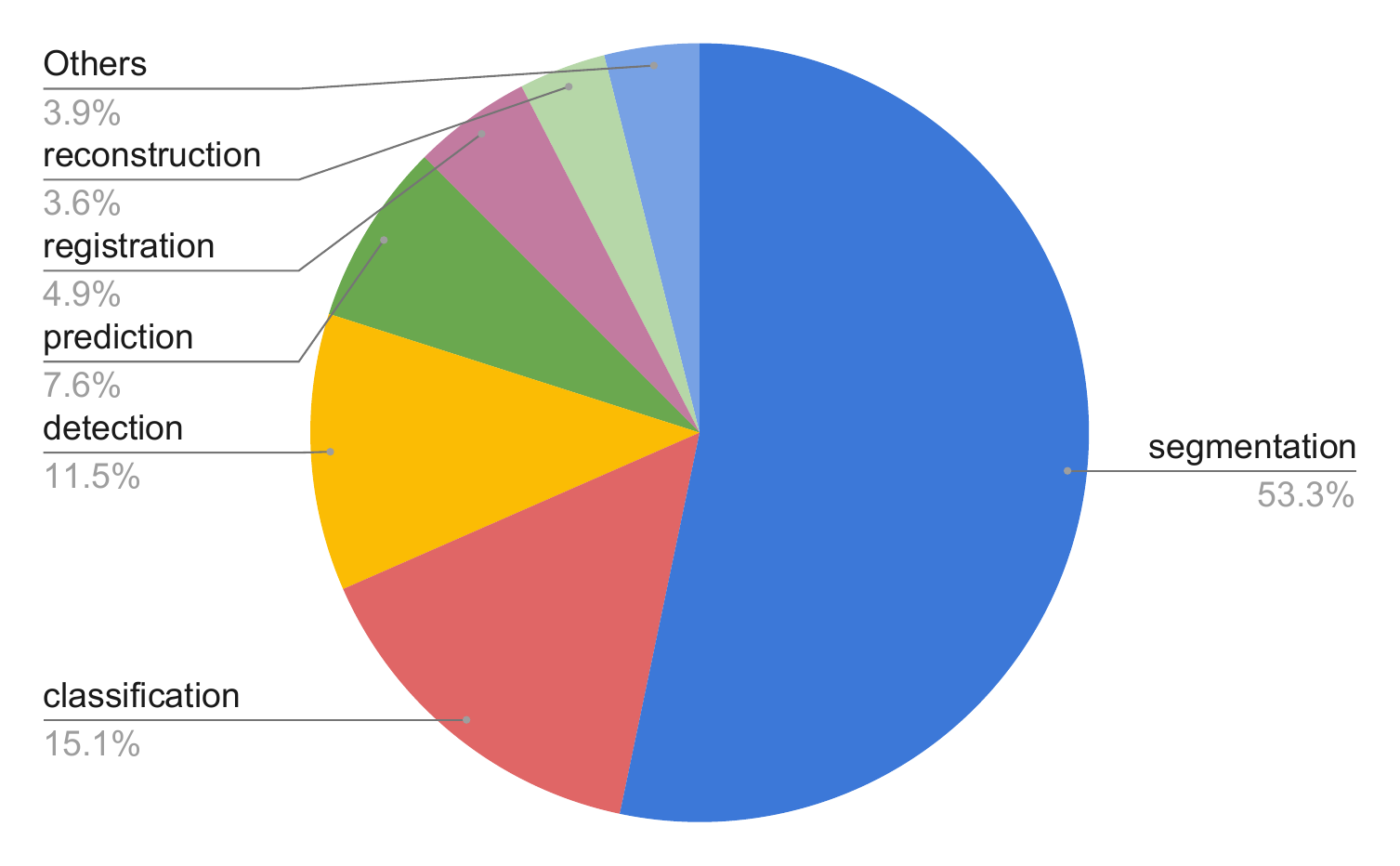}
    \caption{Distribution of different applications with uncertainty estimation in reviewed papers}
    \label{fig: application}
\end{figure}

In this section, we mainly focus on the introduction of recent research that studies the uncertainty in medical image reconstruction, registration, detection, prediction, classification, and segmentation. Figure \ref{fig: application} shows the application types.


Apart from the main applications mentioned above, uncertainties in other medical image tasks such as microstructure estimation \citep{ye2020improved, adler2019uncertainty}, image quality estimation \citep{shaw2020heteroscedastic}, survival analysis \citep{feng2020brain, gomes2021building}, risk analysis \citep{qian2020cq}, image denoising \citep{laves2020uncertainty, laves2020well, cui2022pet}, cellularity assessment \citep{li2022ultra}, lesion localization \cite{wu2021quantifying, duchateau2016infarct, schobs2022uncertainty}, etc, are also mentioned and studied. Since the uncertainty is similar to the methods we introduced before, we will not go into details about those applications.

\subsection{Medical image reconstruction}
Medical image reconstruction plays a critical role in modern healthcare and medical imaging. It involves creating high-quality and accurate images of the internal structures of the human body from acquired raw data. The real-world factor is that medical imaging is subject to various sources of variability, including patient motion, imaging artifacts, and variations in imaging protocols. Therefore, developing reconstruction algorithms that can handle such variability and generalize well across different imaging scenarios is an ongoing challenge. Advanced ML reconstruction algorithms, despite providing good reconstruction performance, often lack reliability and explainability (e.g., understanding why a specific reconstruction was produced or tracing back when the results become unreliable), limiting the adoption and acceptance of these methods in clinical application. Therefore, studying reconstruction uncertainty is of great importance to ensure reconstruction reliability and provide explainable results. Table \ref{tab: reconstruction} shows the related medical image reconstruction methods considering reconstruction uncertainty.
\begin{table*}
  \centering
  \caption{Uncertainty quantification methods in medical image reconstruction}
  \scalebox{1}{
  \begin{tabular}{lllllll}
 \hline
&Publications& Uncertainty  & Number of  & Clinical applications \\
&&  methods&Dataset   & \\
\hline
\multirow{4}{*}{MC methods}& \cite{neumann2014robust}&MCMC& 1&Electromechanical heart MRI reconstruction \\
&\cite{zhou2020bayesian}&MCMC& 1 &PET reconstruction\\
&\cite{luo2023bayesian}& MCMC &1& MRI reconstruction \\
&\cite{edupuganti2020uncertainty} &MCD& 1&knee MRI reconstruction\\
\hline
\multirow{5}{*}{Bayesian inference} &\cite{wallman2014computational}&PD &1 &CT-derived ventricular model reconstruction \\
&\cite{zhang2019reducing} &PD & 1&Knee MRI reconstruction \\
&\cite{narnhofer2021bayesian}&PD & 1&Undersampled MRI reconstruction \\
&\cite{vlavsic2023estimating}&PD& 1& Low/standard-dose PET reconstruction\\
&\cite{barbano2021quantifying}&BNN& 1 &Sparse view CT reconstruction \\
\hline
&\cite{wang2023novel}&Fuzzy sets & 1& COVID-19 CT reconstruction \\
\hline
\end{tabular}
}
\label{tab: reconstruction}
\end{table*}

\subsubsection{Bayesian inference}
Bayesian inference is the most common approach to quantifying reconstruction uncertainty. In 2014, Wallman et al. developed a electrical propagation model based on Bayesian inference with probabilistic distribution for tissue conduction properties inferred from electroanatomical data and designed strategies to optimize the location and number of measurements required to maximize information and reduce uncertainty \citep{wallman2014computational}. The proposed method provides a simultaneous description of clinically relevant electrophysiological conduction properties and their associated uncertainty for various levels of noise. 

In 2019, Zhang et al. proposed an uncertainty reduction model in undersampled MRI reconstruction with an active acquisition that, at inference time, dynamically selects the measurements to take and iteratively refines the prediction to reduce the reconstruction error and uncertainty \citep{zhang2019reducing}. The authors modeled pixel-level uncertainty as a Gaussian distribution centered at reconstruction mean and with variance similar to the method proposed by \citep{kendall2017uncertainties}. 

In 2021, Narnhofer et al. introduced a Bayesian variational framework to quantify the epistemic reconstruction uncertainty \citep{narnhofer2021bayesian}. They first solved the linear inverse problem of undersampled MRI reconstruction in a variational setting and then obtained epistemic uncertainty from a multivariate Gaussian distribution, whose mean and covariance matrix are learned in a stochastic optimal control problem. In the same year, Barbano et al. developed a scalable, data-driven, knowledge-aided computational framework to quantify reconstruction uncertainty via Bayesian neural networks \citep{barbano2021quantifying}. This framework extended to a developed greedy iterative training scheme, deep gradient descent, and recast it within a probabilistic framework. The last layer of each block is Bayesian, with the rest of the layers remaining deterministic to achieve scalability. The framework is showcased on computed tomography with either sparse or limited view data and exhibits competitive performance with respect to SOTA benchmarks, e.g., total variation, deep gradient descent, and learned primal-dual.

In 2023, Vlavsic et al. proposed a DL-based posterior sampling method for uncertainty quantification in PET image reconstruction \citep{vlavsic2023estimating}. The method is based on training a conditional generative adversarial network whose generator approximates sampling from the posterior in Bayesian inversion. The generator is conditioned to a reconstruction from a low-dose PET scan obtained by a conventional reconstruction method. It can, therefore, generate corresponding standard dose PET images.

\subsubsection{Monte Carlo methods}
Monte Carlo methods are also popular for reconstruction uncertainty estimation. In 2014, Neumann et al. presented a stochastic method to estimate the parameters of an image-based electromechanical heart model and the corresponding uncertainty due to measurement noise \citep{neumann2014robust}. First, Bayesian inference was applied to fully estimate the posterior probability density function (PDF) of the model. Second, MCMC sampling was used with computationally tractable designing that employed a fast Polynomial Chaos Expansion-based surrogate model instead of the true forward model. Then, the mean-shift algorithm was used to automatically find the modes of the PDF and select the most likely one while being robust to noise. 

In 2020, Zhou et al. provided a framework for performing infinite-dimensional Bayesian inference and uncertainty quantification for image reconstruction with Poisson data \citep{zhou2020bayesian}. They first introduced a positivity-preserving reparametrization and a dimension-independent MCMC algorithm based on the preconditioned Crank Nicolson Langevin method, in which a primal-dual scheme is used to compute the offset direction. Then, a fusion method that combines the model discrepancy and maximum likelihood estimation was proposed to determine the regularization parameter in the hybrid prior. In the same year, Edupuganti et al. quantified the image recovery uncertainty within DL models \citep{edupuganti2020uncertainty}. First, variational autoencoders (VAEs) were first leveraged to develop a probabilistic reconstruction scheme that maps out (low-quality) short scans with aliasing artifacts to the diagnostic-quality ones and then encoded the acquisition uncertainty in a latent code and naturally offers a posterior of the image from which one can generate pixel variance maps using MCD. 

In 2023, Luo et al. introduced a framework that enables efficient sampling from learned probability distributions for MRI reconstruction where the samples were drawn from the posterior distribution given the measured k-space using MCMC \citep{luo2023bayesian}. Therefore, in addition to the maximum posterior estimate for the image using the log-likelihood, the minimum mean square error estimate and uncertainty maps can also be computed from those drawn samples. 

\subsubsection{Non-probabilistic methods} 
Fuzzy theory can also be applied to quantify image reconstruction uncertainty. In 2023, Wang et al. proposed a new fuzzy metric to characterize image reconstruction uncertainty. It first designed a fuzzy hierarchical fusion attention neural network based on multiscale guided learning \citep{wang2023novel} to convert input images into a fuzzy domain using fuzzy membership functions. The uncertainty of the pixels was processed using the proposed fuzzy rules, and then the output of the fuzzy rule layer was fused with the result of the convolution in the neural network. Simultaneously, a multiscale guided-learning dense residual block and pyramidal hierarchical attention module were designed to extract hierarchical image information. Finally, a recurrent memory module with a residual structure was used to process the output features of the hierarchical attention modules and a recursive sub-pixel reconstruction module was used at the tail of the network to reconstruct the images.
\subsection{Medical image registration}
\begin{table*}
  \centering
  \caption{Uncertainty quantification methods in medical image registration}
  \scalebox{0.9}{
  \begin{tabular}{lllllll}
 \hline
&Publications& Uncertainty  & Number of   &Clinical applications \\
& & methods& dataset   &   \\
\hline
\multirow{3}{*}{MC methods}& \cite{risholm2013bayesian}   &MCMC& 1& Neurosurgery for resection of brain tumor\\
&\cite{le2016quantifying}    &MCMC&1&Medical image registration\\
&\cite{xu2022double} &MCD& 2&  Abdominal CT-MRI registration\\

\hline
\multirow{7}{*}{Bayesian inference}&\cite{oreshkin2013uncertainty} &PD&1& Medical image registration\\
&\cite{parisot2014concurrent} &PD &1 &Atlas to diseased patient registration\\
&\cite{yang2015uncertainty} &  PD&2 &Heart\&brain image registration\\
&\cite{wang2018efficient}   &  PD&2 &Synthetic and brain MRI registration\\
&\cite{khawaled2022npbdreg} & PD&4 &Brain MRI registration&\\
&\cite{wachinger2014gaussian}   &GP &2& MRI image registration\\
&\multirow{2}{*}{\cite{peter2021uncertainty}}&  \multirow{2}{*}{GP}&\multirow{2}{*}{4}& Histology inter-modal, Optical Coherence  \\
&&&& Microscopy image and chest CT scan registration\\
\hline
\multirow{2}{*}{Hybrid methods}& \multirow{2}{*}{\cite{gong2022uncertainty} }  &MCD, Bootstrap& \multirow{2}{*}{2}& \multirow{2}{*}{Deformable medical image registration}\\
 &&Ensembles&\\
\hline

\end{tabular}
}
\label{tab: registration}
\end{table*}
Medical image registration, a fundamental technique in medical image preprocessing, aligns and overlays multiple images of the same patient or anatomical region acquired at different times, from different modalities, or from different imaging devices. By aligning the images, it becomes easier to compare and analyze changes in anatomy or pathology over time. However, given the current SOTA registration technology and the difficulty of the problem, an uncertainty measure that highlights locations where the algorithm had difficulty finding a suitable alignment can be beneficial. According to our literature review, the predominant way to quantify the registration uncertainty is by using summary statistics of the transformation distribution. Table \ref{tab: registration} listed the related papers. For medical image registration tasks, two probabilistic methods, Bayesian inference and MC methods, are mainly developed.


\subsubsection{Bayesian inference}
In 2013, Oreshkin et al. proposed a voxel selection strategy for medical image registration with the uncertainty of the transformation parameters \citep{oreshkin2013uncertainty}. First, a Bayesian framework was used to build a voxel sampling probability field (VSPF) based on the variance of this optimal Bayesian estimator, different voxel subsets were then sampled based on the obtained VSPF. 

In 2014, Parisot et al. presented a graph-based concurrent brain tumor segmentation and atlas to disease patient registration framework based on a unified pairwise discrete Markov Random Field (MRF) model with non-uniform sampling \citep{parisot2014concurrent}, following the sampling method proposed in \citep{oreshkin2013uncertainty}. First, to get an appropriate sampling solution and reduce memory requirements, content-driven samplings of the discrete displacement set and the sparse grid were considered based on the local segmentation and registration uncertainties recovered by the min marginal energies \citep{kohli2008measuring}. Then, both segmentation and registration problems were modeled using a unified pairwise discrete MRF model on a sparse grid superimposed on the image domain. The registration uncertainty is then calculated by normalizing the min-marginals over all the possible displacements associated with the same segmentation label, while the segmentation uncertainty is evaluated by measuring the energy variation when the segmentation label changes. 

In 2015, Yang et al. followed the idea that consisted of mapping displacement into uncertainty by energy information and approximating the covariance matrix by the inverse of the Hessian of the registration energy to quantify registration uncertainty for large deformation diffeomorphic metric mapping \citep{yang2015uncertainty}. The covariance matrix of the Gaussian process posterior distribution was also applied in \citep{wachinger2014gaussian} to estimate registration uncertainty. 

In 2018, Wang et al. presented a large deformation diffeomorphic metric mapping approach similar to \citep{yang2015uncertainty} that models posterior distribution with a Laplace approximation of Bayesian registration models \citep{wang2018efficient}. 

In 2021, Peter et al. introduced a principled strategy for the construction of a gold standard for deformable registration by building on the true transformation into a Gaussian process model and then annotating the most informative location in an active learning fashion to minimize the uncertainty of the true transformation \citep{peter2021uncertainty}. It should be noted that, in addition to a landmark correspondence for each queried location, this framework supports the specification of an annotation uncertainty, either directly estimated by the annotator or obtained by merging annotations from multiple users.

In 2022, khawaled et al. developed a non-parametric Bayesian method to assess the uncertainty in diffeomorphic deformable MRI registration \citep{khawaled2022npbdreg}. It sampled the true posterior distribution of the network weights by noise injection in the training loss gradients with the Adam optimizer and estimated the registration uncertainty according to the voxel-wise diagonal variance. 

\subsubsection{Monte Carlo methods}
In 2013, Risholm et al. proposed a non-rigid registration framework where conventional dissimilarity and regularization energies were included in the likelihood and the prior distribution on deformations respectively through Boltzmann’s distribution \citep{risholm2013bayesian}. MCMC was used to characterize the posterior distribution with Boltzmann temperature hyper-parameters marginalized under broad uninformative hyper-prior distributions, permitting the estimation of the most likely deformation as well as the associated uncertainty. 

In 2016, Le Folgoc et al. investigated uncertainty quantification under a sparse Bayesian model of medical image registration with a focus on the theoretical and empirical quality of uncertainty estimates derived under the approximate scheme and under the exact model \citep{le2016quantifying}. In this paper, the authors implemented an (asymptotically) exact inference scheme based on reversible jump MCMC sampling to characterize the posterior distribution of the transformation. 

In 2022, Xu et al. proposed a mean-teacher registration framework, which incorporates an additional temporal consistency regularization term by encouraging the teacher model’s prediction to be consistent with that of the student model \citep{xu2022double}. Instead of searching for a fixed weight, the teacher model enables automatically adjusting the weights of the spatial regularization and the temporal consistency regularization by taking advantage of the transformation uncertainty and appearance uncertainty calculated based on MCD. 

\subsubsection{Hybrid methods}
Compared with other probabilistic uncertainty quantification methods, model ensemble is less popular and is usually used with other methods to construct a hybrid model. In 2023, Gong et al. \citep{gong2022uncertainty} proposed a predictive module to learn the registration and uncertainty in correspondence simultaneously by inducing three empirical randomness and registration error-based uncertainty prediction methods: MCD, deep ensembles, and Bootstrap. 

In general, the majority of existing research focuses on trying out different summary statistics as well as means to exploit registration uncertainty. Those researches do have promising contributions, e.g., risk assessment based on the trustworthiness of the registered image data. 

\subsection{Medical image detection}
Medical image detection, aiming at detecting small or subtle abnormalities, anatomical structures, lesions, tumors, or other pathologies, plays a vital role in early diagnosis, treatment planning, and medical conditions monitoring. Images used for detection may have low contrast, low signal-to-noise ratio, or be overshadowed by surrounding structures. These factors can make it difficult for detection methods to identify and localize small objects accurately, leading to false negatives or reduced sensitivity. Therefore, it is necessary to estimate the detection uncertainty. Table \ref{tab: registration} listed the related papers. 

\begin{table*}
  \centering
  \caption{Uncertainty quantification methods in medical image detection}
  \scalebox{0.9}{
  \begin{tabular}{lllllll}
 \hline
&Publications& Uncertainty  & Number of   &Clinical applications   \\
& & methods& dataset   &   \\
\hline

\multirow{8}{*}{Bayesian inference}&\cite{kwon2020uncertainty}&PD &2& Retinal blood vessels detection\\
&\cite{araujo2020dr}&PD& 2& Diabetic retinopathy grading diagnosis\\
&\cite{mao2020abnormality}&PD&2&Lung abnormal detection\\
&\cite{akrami2021quantile}&PD &2& Brain lesions detection\\
&\cite{jafari2021u}&PD & 1&Video keyframes detection \\
&\cite{sudarshan2021towards}& PD&1& PET-MRI OoD detection \\
&\cite{wang2022global}&PD&1& Abnormal lymph nodes detection  \\
&\cite{huang2022pixel}&BNN &5&Anomaly detection\\
\hline
\multirow{10}{*}{MC methods}&\cite{leibig2017leveraging}&MCD&1 &Diabetic retinopathy detection\\
&\cite{gill2019uncertainty}&MCD& 1&Focal cortical dysplasia detection  \\
&\cite{nair2020exploring}&MCD&1&Sclerosis lesion detection\\
&\cite{ghoshal2020estimating}&MCD&1& COVID-19 detection \\
&\cite{yang2021exploring}& MCD&1&Lung nodule detection\\
&\cite{dong2021rconet}&MCD  &1& COVID-19 detection \\
&\cite{calderon2021improving} &MCD &1 & Breast cancer detection\\
&\cite{tang2022unified} &MCD&4& Retinal vessel detection\\
&\cite{ghoshal2021cost}&MC sampling& 1& COVID-19 detection \\
&\cite{bhat2021using}&TTD&1&Liver lesions detection\\
\hline
\multirow{2}{*}{TTA}&\cite{ayhan2020expert}&TTA&2& Diagnosing diabetic retinopathy\\
&\cite{ayhan2022test}&TTA&1&Diabetic retinopathy detection\\
\hline
\multirow{2}{*}{DST}&\cite{ben2022fusion}&Basic DST&1& Pneumonia diagnosis\\
&\cite{rahman2023demystifying}&Basic DST&1&Fetal plane detection\\
\hline
\multirow{1}{*}{Model ensemble} &\cite{kabir2022aleatory}&Ensemble &1 & COVID-19 detection\\
\hline
\multirow{1}{*}{Interval analysis}&\cite{mazoure2022dunescan}& Confidence interval & 1& Skin cancer detection\\
\hline
\multirow{7}{*}{Hybrid methods}&\multirow{2}{*}{\cite{tabarisaadi2022uncertainty}}&MCD, Ensemble,&\multirow{2}{*}{1} &\multirow{2}{*}{Skin cancer detection}\\
&&Spectral GP&&\\
&\multirow{2}{*}{\cite{asgharnezhad2022objective}}&MCD, Ensemble, &\multirow{2}{*}{1} &\multirow{2}{*}{COVID-19 detection}\\
&& Ensemble-MCD&&\\
&\cite{javadi2022towards}&TTD, TTA&1& Prostate cancer detection\\
&\cite{abdar2023uncertaintyfusenet}&Ensemble-MCD &2& COVID-19 detection\\
&\multirow{2}{*}{\cite{linmans2023predictive}} &\multirow{2}{*}{MCD, Ensembles}&\multirow{2}{*}{5}& Lymph node tissue, prostate  \\
&&&& cancer/biopsies, foreign tissue detection \\
\hline
\end{tabular}
}
\label{tab: detection}
\end{table*}
\subsubsection{Bayesian inference}
In 2020, Araujo et al. proposed a deep learning-based Diabetic Retinopathy grading computer-aided diagnosis system that supports its decision by providing a medically interpretable explanation and estimation of prediction uncertainty with a novel Gaussian-sampling approach and a multiple-instance learning framework, allowing the ophthalmologist to measure how much that decision should be trusted \citep{araujo2020dr}. In the same year, Mao et al. proposed an abnormality detection approach based on an autoencoder that outputs not only the reconstructed normal version of the input image but also a pixel-wise uncertainty prediction with probabilistic distribution \citep{mao2020abnormality}. 

In 2021, Sudarshan et al. proposed a sinogram-based uncertainty-aware deep BNN framework to estimate a standard-dose PET image \citep{sudarshan2021towards}. Here, the detection uncertainty is modeled through the per-voxel heteroscedasticity of the residuals between the predicted and the high-quality reference images. Jafari et al. presented a video keyframe landmark detection framework by leveraging the uncertainty of landmark prediction obtained from a deep Bayesian network \citep{jafari2021u}. Akrami et al. described a quantile regression VAE model to avoid variance shrinkage problems by estimating conditional quantiles for the given input image \citep{akrami2021quantile}. Using the estimated quantiles, the conditional mean and variance for input images were computed under the Gaussian model to estimate detection uncertainty.

In 2022, Huang et al. presented an uncertainty-aware prototypical transformer model, considering both the anomaly diversity and uncertainty to achieve accurate pixel-level visual anomaly detection \citep{huang2022pixel}. First, a memory-guided prototype learning transformer encoder was designed to learn the diversity of prototypical representations of anomalies. Second, an anomaly detection uncertainty quantizer was designed by a Bayesian Neural Network with Gaussian distribution to learn the distributions of anomaly detection. Then, an uncertainty-aware transformer decoder was proposed to leverage the detection uncertainties to guide the model to focus on the uncertain areas. In the same year, Wang et al. proposed an improved Mask RCNN framework with a global-local channel attention mechanism and multi-task Gaussian inference-based uncertainty loss for the detection of abnormal lymph nodes in MR images \citep{wang2022global}.

\subsubsection{Monte Carlo methods}
In 2017, Leibig et al. evaluated the impact of MCD-based Deep Bayesian uncertainty measures in diagnosing diabetic retinopathy and showed that uncertainty-informed decision referrals could improve diagnostic performance \citep{leibig2017leveraging}. Similar research has been investigated in \citep{gill2019uncertainty}, \citep{ghoshal2020estimating} and \citep{nair2020exploring} for the detection of COVID-19, focal cortical dysplasia detection and lesion, respectively. 

In 2021, Yang et al. improved performance of a detection CNN performance with two different bounding-box-level (or instance-level) uncertainty estimates with predictive variance and MC sampling variance, respectively \citep{yang2021exploring}; Dong et al. proposed a novel deep network for robust COVID-19 detection that employs Deformable Mutual Information Maximization (DeIM), Mixed High-order Moment Feature (MHMF), and Multiexpert Uncertainty-aware Learning (MUL) \citep{dong2021rconet}. With DeIM, the mutual information between input data and the corresponding latent representations can be estimated and maximized to capture compact and disentangled representational characteristics. MHMF is used to extract discriminative features of complex distributions, and MUL creates multiple parallel MCD networks for each image to evaluate uncertainty and thus prevent performance degradation caused by the noise in the data. 

The same year, Ghoshal et al. proposed a Bayesian inference model with MC sampling \citep{ghoshal2021cost} for uncertainty quantification and measured bias-corrected uncertainty using the Jackknife resampling technique \citep{sahinler2007bootstrap}; Bhat et al. proposed to use TTD to reduce false positive detections made by a neural network using an SVM classifier trained with features derived from the uncertainty map of the neural network prediction \citep{bhat2021using}.

Later in this year, Calderón-Ramírez et al. explored the impact of using unlabeled data through the implementation of a recent successful semi-supervised approach, MixMatch \citep{berthelot2019mixmatch}, for breast cancer detection on mammogram images \citep{calderon2021improving}. They improved uncertainty estimations, i.e., Normalized entropy of Softmax, Maximum value of Softmax and MCD, by using unlabeled data under regimes with a very limited number of labeled observations for training. Moreover, following \citep{asgharnezhad2022objective}, the authors used the proposed "uncertainty confusion matrix" that groups uncertainty estimations for each of a model's predictions according to their “correctness” and “confidence.”  Based on this, the authors proposed an uncertainty-balanced accuracy to ease the comparison of uncertainty estimation approaches in real-world usage scenarios. 

\subsubsection{Model ensemble}
In 2022, Kabir et al. proposed an aleatory-aware deep uncertainty quantification method for COVID-19 detection with an application for transfer learning and deep ensembles that converted the outputted K-nearest posteriors of each DNN into opacity scores to represent aleatory uncertainty \citep{kabir2022aleatory}. 

\subsubsection{Non-probabilistic methods}
In 2020, Ayhan et al. studied an intuitive framework based on TTA to quantify the diagnostic uncertainty of a state-of-the-art DNN for diagnosing diabetic retinopathy \citep{ayhan2020expert}. Based on the first work, Ayhan et al. proposed a simple but effective method using traditional data augmentation methods such as geometric and color transformations at test time, allowing us to examine how much the network output varies in the vicinity of examples in the input spaces \citep{ayhan2022test}.

In 2022, Ben et al. proposed a disease dection approach based on a DST-based evidence fusion theory, allowing the combination of a set of deep learning classifiers to provide more accurate disease detection results \citep{ben2022fusion}. The main contribution of this work is the application of Dempster's rule for the fusion of five pre-trained convolutional neural networks (CNNs) including VGG16, Xception, InceptionV3, ResNet50, and DenseNet201 for the diagnosis of pneumonia from chest X-ray images. In the same year, Mazoure et al. released a web server, Deep Uncertainty Estimation for Skin Cancer (DUESC) \citep{mazoure2022dunescan}, that performs an intuitive, in-depth analysis of uncertainty in commonly used skin cancer classification models based on CNNs and confidence intervals.

\subsubsection{Hybrid methods}
In 2021, Javadi et al. proposed a UNet-based deep network for prostate cancer detection in systematic biopsy considering both the label and model uncertainty using TTA and TTD, respectively \citep{javadi2022towards}. Uncertainty metrics were then used to report the cancer probability for regions with high confidence to help the clinical decision-making during the biopsy procedure.

 In 2022, Tabarisaadi et al. studied the automatic diagnosis of skin cancer using dermatologist spot images \citep{tabarisaadi2022uncertainty}. Three different uncertainty-aware training algorithms (MCD, Model ensembling, and Spectral Normalized Neural Gaussian Process \citep{liu2020simple}) were utilized to detect skin cancer. In the same year, Asgharnezhad et al. applied and evaluated three uncertainty quantification techniques, MCD, Ensembles and Ensembles-MCD, for COVID-19 detection \citep{asgharnezhad2022objective}. Moreover, a novel concept of uncertainty confusion matrix was proposed and new performance metrics for the objective evaluation of uncertainty estimates were introduced. 

 In 2023, Abdar et al. presented a simple but efficient deep learning feature fusion model, UncertaintyfuseNet, for COVID-19 detection by using the Ensemble-MCD technique to model detention uncertainty and the obtained results prove the efficiency of the model with robustness to noise and unseen data \citep{abdar2023uncertaintyfusenet}. In the same year, Linmans et al. provided a benchmark for evaluating prevalent uncertainty methods by comparing the uncertainty estimates on both ID and realistic near and far OoD data on a whole-slide level using MCD and model ensembles \citep{linmans2023predictive}. 
\subsection{Medical image prediction}
Radiomics aim to predict future outcomes or conditions from medical images. Although it has been widely studied recently, it also has certain limitations. For example, disease progression in many medical conditions is complex and multifactorial. Predicting the progression or response to treatment from medical images alone may oversimplify the underlying dynamics. Moreover, radiomic methods often encounter uncertainty and variability in image-based measurements. Quantifying and addressing these uncertainties is crucial for reliable predictions and their subsequent use in clinical decision-making. Table \ref{tab: prediction} lists the related work. 

\begin{table*}
  \centering
  \caption{Uncertainty quantification methods in medical image prediction}
  \scalebox{0.85}{
  \begin{tabular}{lllllll}
 \hline
&Publications& Uncertainty  & Number of  &Clinical   \\
&&  methods& Dataset   &application   \\
\hline

\multirow{3}{*}{Bayesian inference}&\cite{bliesener2019efficient}& PD &1&Brain tumor longitudinal monitoring\\ 
&\cite{corrado2020quantifying}&PD&1& Left atrium electro-physiology simulation prediction  \\

&\cite{wu2021quantifying}& Sparse GP&2&  Bone age prediction and lesion localization \\
\hline

\multirow{8}{*}{MC methods}&\cite{rafael2022prediction}&MC sampling &1&  Lung tumour growth prediction \\
&\cite{corrado2023quantifying}&MC sampling &1& Left atrium anatomy prediction\\
&\multirow{2}{*}{\cite{huang2020edge}}&\multirow{2}{*}{MCD}&\multirow{2}{*}{4}& Autism spectrum disorder, Alzheimer, \\
&&&&and ocular diseases prediction\\

&\cite{hemsley2020deep}&MCD&1& Brain metastases/glioblastoma radiation treatment prediction\\
&\cite{kannan2021leveraging}&MCD&1&Assessment of paediatric dysplasia of the hip\\
&\multirow{2}{*}{\cite{dolezal2022uncertainty}}&\multirow{2}{*}{MCD}&\multirow{2}{*}{2}& lung adenocarcinoma and squamous cell\\
&&&&carcinoma prediction\\
\hline

\multirow{6}{*}{DST} &\cite{lian2016robust}&ECM& 2& Lung and esophageal cancer treatment outcomes prediction\\
&\multirow{2}{*}{\cite{lian2016selecting}}&\multirow{2}{*}{ECM}&\multirow{2}{*}{3}& Lung, lymph and esophageal cancer \\
&&&&treatment outcomes prediction\\
&\cite{wu2018treatment}&Basic DST& 4&Cancer treatment outcome prediction\\
&\cite{liu2023evidence}&Basic DST&1& Knee replacement prediction\\
&\cite{ahmad2023prognosis}&Basic DST&2& COVID-19 progression and prognosis prediction\\
\hline

\multirow{2}{*}{Hybrid methods} &\multirow{2}{*}{\cite{jensen2019improving}}& Ensemble, TTA & \multirow{2}{*}{1}& \multirow{2}{*}{Skin conditions prediction}\\
&&MC sampling, MCD&&\\
\hline
\end{tabular}
}
\label{tab: prediction}
\end{table*}

\subsubsection{Bayesian inference}
In 2020, Corrado et al. used a Bayesian probabilistic approach to detect the left atrium derived from cardiac MRI and to quantify the uncertainty about the shape \citep{corrado2020quantifying}. In 2021, Wu et al. proposed an uncertainty-aware deep kernel learning model that permits the estimation of the uncertainty in the prediction by a pipeline of a CNN and a sparse Gaussian Process \citep{wu2021quantifying}. In 2022, Rafael et al. proposed a deep hierarchical generative and Bayesian probabilistic network that, given an initial image of the nodule, predicts whether it will grow, quantifies its future size and provides its expected semantic appearance at a future time and estimates the uncertainty in the predictions from the intrinsic noise in medical images and the inter-observer variability in the annotations \citep{rafael2022prediction}. In 2023, Corrado et al. described the left atrium anatomy using a Bayesian shape model that captures anatomical uncertainty in medical images and validated the model on independent clinical images \citep{corrado2023quantifying}.

\subsubsection{Monte Carlo methods}
In 2018, Jungo et al. proposed an MCD-based full-resolution residual CNN for brain tumor segmentation and survival prediction\citep{jungo2018towards}. In 2019, Bliesener et al. used a neural network to estimate the approximate joint posterior distribution of tracer-kinetic parameters, where uncertainties are estimated for each voxel and are specific to the patient, exam, and lesion \citep{bliesener2019efficient}. The predicted parameter ranges correlate well with tracer-kinetic parameter ranges observed across different noise realizations and regression algorithms. 

In 2020, Huang et al. proposed a concept of MC edge dropout to estimate the predictive uncertainty related to the graph topology \citep{huang2020edge}. After that, Hemsley et al. proposed an MCD-based conditional generative adversarial model for brain metastases or glioblastoma radiation treatment prediction  \citep{hemsley2020deep} and Dolezal et al. trained Bayesian Neural models with MCD to identify lung adenocarcinoma and squamous cell carcinoma \citep{dolezal2022uncertainty}.

\subsubsection{Non-probabilistic methods}
For medical image prediction tasks, DST is the most commonly used non-probabilistic uncertainty quantification method. In 2016, Lian et al. proposed a radiomics feature-based radiotherapy treatment outcomes prediction system using a feature selection method based on DST for modeling and reasoning with uncertain and/or imprecise information \citep{lian2016robust}. The proposed method aimed to reduce the imprecision and overlaps between different classes in the selected feature subspace, thus finally improving the prediction accuracy. Based on the proposed feature selection model, Lian et al. proposed a radiotherapy treatment outcomes prediction system that uses EKNN for radiomic features selection with the consideration of a data balancing procedure and specified prior knowledge \citep{lian2016selecting}. After that, Wu et al. proposed a similar method for cancer treatment outcome prediction with a feature selection module and an EKNN classifier \citep{wu2018treatment}.


In 2023, Ahmad et al. presented a complete COVID-19 progression and prognosis prediction framework using a two-stage reasoning process based on the DST \citep{ghesu2021quantifying}. In the same year, Liu et al. proposed an evidence-aware multi-modal data fusion framework based on DST that considers the reliability of each source data and the prediction output when making a final decision \citep{liu2023evidence}. The backbone models contain an image, a non-image branch and a fusion branch. For each branch, there is an evidence network that takes the extracted features as input and outputs an evidence score, which is designed to represent the reliability of the output from the current branch. The output probabilities along with the evidence scores from multiple branches are combined with Dempster's combination rule to make a final prediction.

\subsubsection{Hybrid methods}
In 2019, Jensen et al. experimentally showed that models trained to predict skin conditions become overconfident and then proposed to train models with a label sampling scheme that takes advantage of inter-rater variability to achieve a better-calibrated model \citep{jensen2019improving}. Thus, Model Ensemble, TTA, MC Batch Normalization \citep{teye2018bayesian} and MCD were used to quantify prediction uncertainty. 

\subsection{Medical image classification}
\begin{table*}
  \centering
  \caption{Uncertainty quantification methods in medical image classification}
  \scalebox{0.85}{
  \begin{tabular}{lllllll}
 \hline
&Publications& Uncertainty  & Number of  & Clinical applications  \\
&&  methods& Dataset   &    \\
\hline

\multirow{10}{*}{DST}
&\cite{tardy2019uncertainty}&SL&2& Mammograms classification\\ 
 &\cite{ghesu2019quantifying}&SL&2&Chest radiograph assessment\\

&\multirow{2}{*}{\cite{yuan2020evidential}}&\multirow{2}{*}{ENN}&\multirow{2}{*}{2}&Breast infiltrating ductal carcinoma and chest\\
&&&&radiograph pneumonia classification\\
&\cite{huang2021covid} &ENN&1&COVID-19 classification\\
&\cite{ghesu2021quantifying} &SL&2 &Chest radiographs abnormalities classification \\
&\cite{xu2022deep}&SL&2&Pancreatic tumor subtype and grade classification\\
&\multirow{2}{*}{\cite{liu2023classifier}}&DST with new basic&\multirow{2}{*}{1}&\multirow{2}{*}{Grading of breast cancer}\\
&& probability assignment&&\\
\hline

\multirow{11}{*}{MC methods} &\multirow{2}{*}{\cite{abdar2021barf}}&\multirow{2}{*}{MCD}&\multirow{2}{*}{4}&COVID-19, chest, optical coherence tomography,\\
&&&&and skin cancer classification\\
&\multirow{2}{*}{\cite{ju2022improving}} &\multirow{2}{*}{MCD}&\multirow{2}{*}{3}&Skin lesions, prostate cancer\\
&&&& and retinal diseases classification\\
&\cite{valen2022quantifying}&MCD&2& Chest and skin cancer classification\\
&\multirow{2}{*}{\cite{feng2022penalized}} &\multirow{2}{*}{MCD}&\multirow{2}{*}{3}& Optical coherence tomography \\
&&&& and chest classification\\
&\cite{ahsan2022active}&MCD&1&Diabetic retinopathy classification\\
&\cite{abdar2022hercules}&MCD&3&Retinal OCT, lung and chest classification\\

&\cite{aljuhani2022uncertainty}&MCD&1&Tumor region classification\\
&\cite{ghoshal2022leveraging}&MC sampling&2&Pancreatic adenocarcinoma grading \\
\hline

\multirow{4}{*}{Bayesian inference} 
&\cite{peressutti2013novel} &PD&4&Cardiac interventions\\
&\cite{thiagarajan2021explanation}&BNN&1&Breast histopathology images classification\\
&\cite{belharbi2021deep}&PD&2&Histology images classification\\
&\cite{liu2022handling}&PD&2& Skin lesion and thorax disease classification\\
&\cite{jimenez2022curriculum} &PD&1&Femur fracture classification\\

\hline
\multirow{4}{*}{Ensemble} &\cite{senousy2021mcua}&Ensemble&1&Breast cancer classification\\
&\multirow{2}{*}{\cite{qendro2021early}}&\multirow{2}{*}{Early exit ensemble}&\multirow{2}{*}{3}& Heart attack, epileptic seizure \\
&&&& and skin melanoma classification \\
&\cite{arco2023uncertainty}&Ensemble&1& Bacterial/viral pneumonia classification\\
\hline

\multirow{2}{*}{Fuzzy sets}&\cite{pham2014nonstationary}&Fuzzy sets &1& Hernia mesh classification\\
&\cite{rahman2023framework} &Fuzzy sets&1&Brain tumour classification \\
\hline

\multirow{2}{*}{Others}&\cite{galdran2019uncertainty}& Soft label&5&Retinal images classification\\
&\cite{del2023labeling}& Soft label&1&Histology image classification\\
\hline

\multirow{9}{*}{Hybrid methods}&\cite{carneiro2020deep}&PD, TTA&1& Polyp classification\\
&\multirow{2}{*}{\cite{abdar2021uncertainty}}&MCD, Ensemble&\multirow{2}{*}{2}&\multirow{2}{*}{Skin cancer classification}\\
&& Ensemble-MCD&&\\
&\cite{gour2022uncertainty}&MCD, CI&3&Breast histopathology images classification\\
&\multirow{2}{*}{\cite{yang2021uncertainty}}&MCD, Ensemble,&\multirow{2}{*}{2}&\multirow{2}{*}{COVID-19 and breast tumor classification}\\
&& and Ensemble-MCD&&\\
&\cite{dawood2023uncertainty} &PD, TTA &2&Cardiac classification\\ 
&\cite{cifci2023deep}&TTD&1&Lung cancer diagnosis and treatment decisions\\
&\cite{mehta2023evaluating}&Ensemble-MCD&3& Skin lesion classification\\
&\multirow{2}{*}{\cite{hamedani2023breast}}&MCD, Ensemble,& \multirow{2}{*}{1}&\multirow{2}{*}{Breast cancer classification}\\
&& and Ensemble-MCD&&\\

\hline
\end{tabular}
}
\label{tab: classification}
\end{table*}

Similar to previous MIA tasks, the performance of medical image classification methods depends on the quality of the image itself and the corresponding annotations. Quantifying instance-level uncertainty helps to classify images where the classification model might be uncertain or incorrect, allowing for manual correction or expert review and improving diagnosis quality and treatment planning. Considering that we have already introduced the main uncertainty quantification methods in sections \ref{subsec: pro} and \ref{subsec: nonpro}, and also the research focused on image classification is similar to the medical image analysis tasks mentioned earlier, here we only briefly describe their corresponding methods, datasets and clinical applications in Table \ref{tab: classification}. 

\subsection{Medical image segmentation}
Medical image segmentation is more challenging than classification tasks due to the inherent variations in the appearance of anatomical structures, leading to potential errors or inaccuracies in defining boundaries or segment structures. Therefore, quantifying pixel/voxel-level uncertainty helps identify regions where the model might be uncertain or incorrect, allowing for manual correction or expert review and improving radiotherapy treatment performance. Table \ref{tab: MC sampling}, \ref{tab: Bayesian inference} and \ref{tab: ensemble} list three main probabilistic uncertainty quantification methods used in medical image segmentation tasks. Table \ref{tab: dst} shows the most frequent non-probabilistic uncertainty methods DST and Table \ref{tab: others-non} shows the rest of the non-probabilistic uncertainty methods. Table \ref{tab: hybrids} shows the hybrid uncertainty quantification methods. Among the retrieved methods, MCD and ENN is the most commonly used probabilistic and non-probabilistic uncertainty quantification method for medical image segmentation, respectively.

In MIA tasks, fully supervised learning has gained huge success based on the satisfying condition that large-scale annotated training datasets are available \citep{ronnebergerconvolutional, myronenko20183d, isensee2018nnu}. However, region labeling in medical image segmentation tasks requires skilled expertise with domain knowledge and careful delineation of boundaries. The contradiction between the increasing demand for segmentation accuracy on the one hand, and the shortage of perfect (precise and reliable) annotations on the other hand has so far limited the performance of learning-based medical image segmentation methods. Therefore, in this section, we focus our uncertainty analysis on semi-supervised medical image segmentation.

Techniques for semi-supervised medical image segmentation can be divided into three groups: graph-constrained methods \citep{xu2016deep, reiss2022graph}, self-learning methods \citep{li2019transformation, min2019two}, and generative adversarial learning methods \citep{mondal2018few, sun2019parasitic}. Though these methods can break the dependence of machine learning models on training labels and the experimental results are promising, the uncertainty caused by the low quality of the images and the lack of annotations still need to be further studied for a more accurate and reliable medical image segmentation model. 

According to our literature review, current uncertainty-based semi-supervised learning methods (the methods be marked in blue color in Tables \ref{tab: MC sampling}, \ref{tab: Bayesian inference}, \ref{tab: ensemble}, \ref{tab: dst}, \ref{tab: others-non} and \ref{tab: hybrids}) can be classified into two main groups: consistency learning \citep{yu2019uncertainty, shi2021inconsistency} and uncertainty-aware learning \citep{sedai2019uncertainty, meyer2021uncertainty}. Consistent learning regularizes the model's predictions to be consistent across different perturbations of the same input and imposes feature-level, data-level, model-level, or task-level consistency on unlabeled data. The common applications are to optimize a teacher-student or multi-view framework with consistent learning, where the teacher/main model provides consistent predictions for guiding the student/ auxiliary model. Uncertainty-aware learning integrates estimated uncertainty into the training process directly when dealing with a mix of labeled and unlabeled data. It leverages the unlabeled data to enhance the model's predictions while providing uncertainty estimates reflecting the model's confidence in those predictions. 

In the rest of the section, we will introduce the semi-supervised medical image segmentation methods with uncertainty quantification in detail.  
\begin{table*}
  \centering
  \caption{Bayesian inference-based uncertainty quantification for medical image segmentation. The semi-supervised methods are highlighted in blue.}
  \scalebox{0.9}{
  \begin{tabular}{lllllll}
 \hline
Publications& Uncertainty  & Number of   &Clinical applications  \\
&  methods& Dataset  &  \\
\hline

\cite{parisot2014concurrent}&PD &2&Low-grade glioma and brain tumor segmentation\\

\cite{le2016sampling}&PD&1& Brain tumor segmentation\\

\cite{ghoshal2019estimating}&PD&1& Nuclei images segmentation\\
\cite{wang2018interactive}&PD&2& Organs and brain tumor core segmentation\\
\cite{behnami2019dual}&PD&1& Infants born MRI tumor segmentation \\
\cite{ouyang2019weakly}& PD&1&Pneumothorax segmentation\\
\cite{baumgartner2019phiseg} &Hierarchical PD &2& thoracic and prostate segmentation \\
\cite{camarasa2021quantitative}& PD&1&Carotid artery segmentation\\
\textcolor{blue}{\cite{luo2021efficient}}&PD &1& Nasopharyngeal carcinoma segmentation \\
\cite{zhang2021multi}& PD &1&Liver tumor segmentation\\
\cite{zhao2021umrn}&PD&1&Carotid artery segmentation\\
\cite{li2022learning}&PD  &2 &Subcortical structures segmentation\\
\cite{li2021hematoma}& Multi-head PD &1 &Intracranial hemorrhage segmentation\\
\cite{luo2021efficient} &PD &1&Nasopharyngeal carcinoma segmentation\\

\cite{mahani2022bounding}&PD&1& Skin lesions segmentation\\
\cite{wang2022uncertainty} &PD&2&Cardiac and skin lesion segmentation \\
\cite{liu2022ahu} & PD &2&Atrial, brain tumor, liver tumor segmentation\\
\cite{xie2022uncertainty}&PD with confidence map &3&Ultrasound Image segmentation\\
\cite{li2022region}&PD &1&Brain tumor segmentation\\
\cite{diao2022unified}&PD &4&Soft tissue, lymphoma and liver tumor segmentation\\
\cite{jones2022direct}&PD &1&Brain tumor segmentation, tissue class prediction\\
\multirow{2}{*}{\textcolor{blue}{\cite{luo2022semi}}} &\multirow{2}{*}{PD}&\multirow{2}{*}{3}&Nasopharyngeal carcinoma, brain tumor \\ 
&&&and pancreas segmentation\\
\textcolor{blue}{\cite{shi2023uncertainty}}&PD&2&Neck tumor segmentation\\
\cite{zhang2023uncertainty} &PD&2&Atrial segmentation, brain tumor segmentation\\
\cite{islam2023paced}&PD&2&Breast segmentation\\
\hline
\textcolor{blue}{\cite{sedai2019uncertainty}}&BNN&1&Optical coherence tomography segmentation\\
\textcolor{blue}{\cite{xia2020semia}}&BNN&2&Pancreas and liver tumor segmentation\\
\cite{bian2020uncertainty}&BNN &2&Retinal OCT images segmentation\\
\cite{kwon2020uncertainty}&BNN &2&Ischemic stroke lesion segmentation, blood vessels detection\\
\cite{senapati2020bayesian}&BNN&1& Liver segmentation and disease classification \\
\cite{krygier2021quantifying}&BNN&2&Spine and aorta segmentation\\
\cite{li2021uncertainty}&BNN&2&Lung and nasal endoscopy segmentation\\
\hline
\end{tabular}
}
\label{tab: Bayesian inference}
\end{table*}
\begin{table*}
  \centering
  \caption{MC methods-based uncertainty quantification for medical image segmentation. The semi-supervised methods are highlighted in blue.}
  \scalebox{0.9}{
  \begin{tabular}{llllllll}
 \hline
Publication& Uncertainty  & Number of  &Clinical applications \\
&  methods& dataset   & \\
\hline
\cite{jungo2018effect} &MCD&2&Synthetic and brain tumor segmentation \\
\cite{jungo2018towards}&MCD&1&Brain tumor Segmentation, Survival Prediction\\
\cite{seebock2019exploiting}&MCD&6& Retinal OCT anatomy segmentation \\
\cite{hu2019supervised}&MCD&2&Lung nodule and prostate segmentation\\
\textcolor{blue}{\cite{yu2019uncertainty}}&MCD&1&Left atrium segmentation\\
\textcolor{blue}{\cite{soberanis2020uncertainty}}&MCD&2&Pancreas and spleen segmentation\\
\textcolor{blue}{\cite{wang2020double}}&MCD&2& Left atrium and kidney segmentation\\
\textcolor{blue}{\cite{xia2020uncertainty}} &MCD&4&Pancreas segmentation \\
\cite{monteiro2020stochastic}&MCD&2&Thorax and brain tumor segmentation \\
\cite{ruan2020mt}&MCD&1&Renal tumors segmentation \\
\cite{liu2020exploring}&MCD&1&Prostate zonal segmentation\\
\cite{hu2020coarse}&MCD&1&Natural killer T cell and lymphoma segmentation\\
\cite{nair2020exploring}&MCD&1& Sclerosis lesion detection and segmentation\\
\cite{wickstrom2020uncertainty}&MCD &1& Polyp segmentation\\
\cite{hasan2021multi}&MCD &1&Cardiac segmentation\\
\textcolor{blue}{\cite{meyer2021uncertainty}}&MCD&3&Prostate zones segmentation\\
\cite{wu2021uncertainty}&MCD&2&Mitochondria segmentation\\
\cite{cao2021dilated}&MCD&1& Breast segmentation\\
\textcolor{blue}{\cite{wang2021tripled}} &MCD&2& Cardiac and prostate segmentation\\
\cite{ghoshal2021estimating} &MCD&2& Cell and brain tumor detection\\
\cite{rousseau2021post} &MCD&2& Ischemic stroke and brain tumor segmentation \\
\cite{balagopal2021deep}&MCD&1&post-operative prostate cancer radiotherapy\\
\cite{wang2021medical}&MCD &3&Thoracic, white matter and skin lesion segmentation\\
\cite{silva2021using} &MCD&4&Brain growth, brain tumor, kidney and prostate segmentation\\
\multirow{2}{*}{\cite{wang2023medical}} &\multirow{2}{*}{MCD}&\multirow{2}{*}{3}&Thoracic skin lesion and brain’s white matter\\
&&&tissue myelination process\\
\textcolor{blue}{\cite{hu2022semi}}&MCD&2&Nasopharyngeal carcinoma segmentation\\
\textcolor{blue}{\cite{qiao2022semi}}&MCD&3&Chest segmentation\\
\textcolor{blue}{\cite{wang2022uncertainty}}&MCD&1&Cardiac segmentation\\
\cite{mojiri2022deep}&MCD&2&white matter hyperintensity segmentation\\
\cite{kuang2022uncertainty}&MCD&1&Perihematomal edema segmentation\\
\cite{tang2022unified} &MCD&4& Nasopharyngeal carcinoma, lung, optic disc segmentation\\
\cite{judge2022crisp}& MCD&3&Cardiac ultrasound, myocardial infarction and lung segmentation\\
\textcolor{blue}{\cite{wang2022semi}} &MCD&2&Cardiac and prostate segmentation\\
\textcolor{blue}{\cite{xiao2022efficient}} &MCD&1&Cardiac segmentation\\
\textcolor{blue}{\cite{zheng2022uncertainty}} &MCD&3&Cardiac, spinal cord gray matter and spleen segmentation \\
\textcolor{blue}{\cite{xiang2022fussnet}}&MCD&2&Left atrium and pancreas segmentation \\
\cite{sambyal2022towards}&MCD&1& Brain tumor segmentation\\
\textcolor{blue}{\cite{lu2023uncertainty}} &MCD &1&Atrial Segmentation\\
\textcolor{blue}{\cite{farooq2023residual}}&MCD&2&Breast masses segmentation\\
\cite{zimmer2023placenta}&MCD&1&Placenta segmentation\\
\hline

\cite{norouzi2019exploiting}&MC sampling&1&Cardiac segmentation\\
\cite{eaton2019easy}&MC sampling&2&white-matter hyperintensity segmentation\\
\cite{huang2020heterogeneity}&MC sampling  &1&Atria and ventricles segmentation \\
\cite{alonso2021use}&MC sampling&1&Retinal OCT images segmentation\\
\cite{zhao2022efficient}&MC sampling&2&Cardiac segmentation\\
\cite{chlebus2022robust}&MC sampling&5&Liver segmentation\\
\textcolor{blue}{\cite{chen2022uncertainty}} &MC sampling &3&Cardiac, spinal cord gray matter and spleen segmentation\\
\textcolor{blue}{\cite{wang2023multi}}&MC sampling&1&Dental panoramic caries segmentation\\
\cite{arega2023automatic}&MC sampling&2& Cardiac pathologies \\
\hline
\cite{natekar2020demystifying}&TTD&1&Brain tumor segmentation\\
\cite{redekop2021uncertainty}&TTD&2&Skin lesion and liver segmentation\\
\textcolor{blue}{\cite{xu2023dual}}&TTD&2 &Brain tumor and left atrial segmentation\\
\hline
\cite{awate2019estimating}&MCMC&4&Brain MRI segmentation\\
\hline

\end{tabular}
}
\label{tab: MC sampling}
\end{table*}
\begin{table*}
  \centering
  \caption{Model Ensemble uncertainty quantification for medical image segmentation. The semi-supervised methods are highlighted in blue.}
  \scalebox{1}{
  \begin{tabular}{lllllll}
 \hline
Publications& Uncertainty  & Number of  &Clinical applications \\
&  method& dataset  &  \\
\hline
\cite{nath2020diminishing} &Ensemble&2&Pancreas and tumor segmentation\\
\cite{fuchs2021practical}&Ensemble&1 & Brain tumor segmentation\\
\textcolor{blue}{\cite{cao2020uncertainty}}& Ensemble&1& Breast mass segmentation\\
\textcolor{blue}{\cite{li2021dual}}& Ensemble&1&COVID-19 lesion segmentation\\
\cite{kushibar2022layer}&Ensemble&2& Breast cancer and cardiac segmentation\\
\cite{guo2022cardiac}&Ensemble&4&Cardiac segmentation\\
\cite{buddenkotte2023calibrating} &Ensemble &2&Cancer and kidney tumor segmentation\\
\cite{zhang2023multi}&Ensemble&2&Tumor segmentation\\
\hline
\end{tabular}
}
\label{tab: ensemble}
\end{table*}

\subsubsection{Bayesian inference}
\paragraph{Consistent learning} In 2021, Shi et al. presented a conservative radical network with probabilistic uncertainty estimation for medical image segmentation \citep{shi2021inconsistency}. The general idea is that if the segmentation result of a pixel becomes inconsistent, this pixel shows a relative uncertainty with probabilistic distribution.

In 2023, Shi et al. proposed an uncertainty-weighted prediction consistency training strategy and a relation-driven consistency training strategy in a semi-supervised fashion for nasopharyngeal carcinoma segmentation \citep{shi2023uncertainty}. The architecture was composed of a shared encoder, a main decoder, and several auxiliary decoders. Various perturbations were applied to the shared encoder’s output to leverage the unlabeled data and enforce consistency between the predictions of the main and auxiliary decoders and uncertainty estimation was applied to avoid being misled by unreliable outputs during training due to annotation scarcity. 
\begin{table*}
  \centering
  \caption{DST-based uncertainty quantification for medical image segmentation. The semi-supervised methods are highlighted in blue.}
  \scalebox{1}{
  \begin{tabular}{lllllll}
 \hline
Publications& Uncertainty  & Number of  &Clinical applications \\
&  methods& dataset   &  \\
\hline
\cite{ghasemi2013novel}&Basic DST&2&Brain MRI segmentation\\

\cite{lelandais2014fusion}&ECM&1&Tumor estimation and dose planning\\
\cite{makni2014introducing}&ECM&1&Prostate multi-parametric segmentation \\
\cite{liu2015new}&DST with fuzzy c-means&1&Brain MRI segmentation\\
\cite{derraz2015joint}&DST optimization&1& Non-small cell lung cancer segmentation \\

\cite{xiao2017vascular}&GD with Dempster's rule &1&Vascular segmentation\\

\cite{lian2017tumor}&ECM&1&Tumor delineation\\
\cite{lian2017accurate}&ECM&1&Tumor Segmentation\\
\cite{lian2017spatial}&ECM&1& Lung cancer Segmentation \\
\cite{lian2018joint}&ECM &1& Lung cancer Segmentation\\
\cite{tavakoli2018brain}&DST with fuzzy c-means&1&Brain MRI segmentation\\
\cite{lima2019modified}&DST with fuzzy c-means&1&Brain MRI segmentation\\
\cite{huang2021belief}&ENN&1&Brain tumor segmentation\\
\citep{huang2021evidential}&ENN&1&Lymphoma segmentation\\

\cite{huang2022evidence}&ENN&1&Brain tumor Segmentation\\
\multirow{2}{*}{\cite{huang2022lymphoma}}&ENN, DST with &\multirow{2}{*}{1}&\multirow{2}{*}{Lymphoma segmentation}\\
&Radial basis function&&\\
\multirow{2}{*}{\cite{fidon2022dempster}}& DST with new&\multirow{2}{*}{1}&\multirow{2}{*}{fetal brain MRI segmentation}\\
&basic probability assignment&&\\
\cite{hu2023trustworthy}&SL&1&Liver tumor segmentation\\
\cite{zou2023evidencecap}&SL&3&Skin lesion, liver and brain tumor segmentation\\
\multirow{2}{*}{\textcolor{blue}{\cite{zhang2023deep}}}&DST with deep &\multirow{2}{*}{4}&\multirow{2}{*}{Brain MRI segmentation}\\
&hyperspherical clustering&&\\
\textcolor{blue}{\cite{huang2023semi}}&ENN&1&Brain tumor segmentation\\
\hline
\end{tabular}
}
\label{tab: dst}
\end{table*}
\paragraph{Uncertainty-aware learning} In 2021, Meyer et al. proposed an uncertainty-aware temporal self-learning (UATS) model to combine the techniques of temporal ensembling and uncertainty-guided self-learning to benefit from unlabeled images \citep{meyer2021uncertainty}. In the same year, Luo et al. proposed a semi-supervised medical image segmentation framework with uncertainty rectified pyramid consistency regularization in \citep{luo2021efficient, luo2022semi}, where uncertainty is estimated via the KL-divergence among multi-scale predictions, which only need a single forward pass compared with MCD.

In 2022, Qiao et al. used a complementary uncertainty pairing rule to dilute the unreliability in semi-supervised learning by assembling reliable annotated data into unreliable unannotated data \citep{qiao2022semi}, where a mixed sample data augmentation method was proposed to integrate annotated data into unannotated data for training the model in a low-unreliability manner. In the same year, Wang et al. proposed an uncertainty-guided pixel contrastive learning method \citep{wang2022uncertainty}, where an uncertainty map for unlabeled data was constructed based on the entropy of the average probability distribution by a well-designed consistency learning mechanism, which generates comprehensive predictions for unlabeled data by encouraging consistent network outputs from two different decoders.
\begin{table*}
  \centering
  \caption{Other non-probabilistic uncertainty quantification methods for medical image segmentation. The semi-supervised methods are highlighted in blue.}
  \scalebox{0.9}{
  \begin{tabular}{lllllll}
 \hline
Publications& Uncertainty  & Number of  &Clinical applications \\
&  methods& dataset  &  \\
\hline
\cite{alberts2016uncertainty}&TTA &15&Brain tumor segmentation\\
\cite{wang2019automatic}&TTA&1&Brain tumor segmentation\\
\cite{xu2022polar}&TTA&1&Prostate ultrasound segmentation\\
\cite{wu2023upl}&TTA&1& Fetal brain segmentation\\
\hline
\cite{zheng2020deep}&Fuzzy sets &2&Pancreas segmentation\\
\cite{bertels2021theoretical}&Soft label &4&Lower-left third molar and brain tumor segmentation\\
\textcolor{blue}{\cite{shi2021inconsistency}} &Conservative and Radical Setting &3& Cancreas and endocardium segmentation\\
\textcolor{blue}{\cite{adiga2022leveraging}}&Plausible sets&1&Left atrium segmentation \\
\cite{huang2022trustworthy}&Fuzzy logic theory &3 &Breast segmentation\\
\hline
\end{tabular}
}
\label{tab: others-non}
\end{table*}
\subsubsection{Monte Carlo methods}
\paragraph{Consistent learning}  In 2019, Yu et al. presented a teacher-student-based uncertainty-aware semi-supervised framework for left atrium segmentation \citep{yu2019uncertainty} with an uncertainty-aware scheme that enables the student model to gradually learn from meaningful and reliable targets by exploiting the uncertainty information using MCD. Following the idea that explores uncertainty caused by lack of annotation, researchers optimized or extended semi-supervised or un-supervised MIA models that use the teacher-student framework with the MC methods. For example, Sedai et al. proposed an uncertainty-guided semi-supervised learning network based on a student-teacher framework for medical image segmentation with MCD \citep{sedai2019uncertainty}. 

In 2022, Chen et al. proposed an MC Sampling-based uncertainty teacher-student framework with dense focal loss and deep co-training \citep{chen2022uncertainty}. In the same year, Xiao et al. designed a teacher-student segmentation method through synchronous training and consistent regular constraints by screening uncertainty assessment with MCD during the training process \citep{xiao2022efficient}; Hu et al. proposed a two-stage teacher-student semi-supervised segmentation framework where an MCD-based uncertainty estimation was introduced to guide the student model to gradually learn reliable predictions from the teacher model \citep{hu2022semi}. In 2023, Farooq et al. proposed a residual-attention-based MCD uncertainty-guided mean teacher framework that incorporates the residual and attention blocks \citep{farooq2023residual}.

In addition to using the MC methods in the teacher-student framework, the MC methods are also popular in multi-view frameworks for uncertainty quantification. In 2020, Xia et al. proposed an uncertainty-aware multi-view co-training framework by exploiting the multi-viewpoint consistency of 3D medical images \citep{xia2020semia, xia2020uncertainty}. They applied co-training by enforcing multi-view consistency generated from MCD on unlabeled data, where an uncertainty estimation of each view is utilized to achieve accurate labeling. A similar approach can be found in \citep{wang2023multi}. In the same year, Zhang et al. proposed an MCD uncertainty-guided mutual consistency learning framework to effectively exploit unlabeled data by integrating intra-task consistency learning from up-to-date predictions for self-ensembling and cross-task consistency learning from task-level regularization to exploit geometric shape information \citep{zhang2023uncertainty}.

\paragraph{Uncertainty-aware learning} In 2019, Sedai et al. proposed an uncertainty-guided semi-supervised learning network based on a student-teacher framework for medical image segmentation \citep{sedai2019uncertainty}. First, a teacher segmentation model was trained from the labeled samples using deep learning with MCD to generate soft segmentation labels and uncertainty maps for the unlabeled set. The student model was then updated using the softly segmented samples and the corresponding pixel-wise confidence of the segmentation quality estimated from the uncertainty of the teacher model using a newly designed uncertainty-based loss function. A similar method with an additional learnable uncertainty consistency loss was proposed in \citep{wang2020double}. 

In 2020, Soberanis-Mukul et al. proposed a segmentation refinement method based on MCD uncertainty analysis and graph convolutional networks \citep{soberanis2020uncertainty}. 

In 2022, Zheng et al. proposed an uncertainty-aware scheme to make models learn segmentation regions purposefully \citep{zheng2022uncertainty}. The model employed MCD as an estimation method to attain uncertainty maps, which serve as a weight for losses to force the models to focus on the valuable region according to the characteristics of supervised learning and unsupervised learning. 

\subsubsection{Model ensemble}

\paragraph{Consistent learning} In 2021, Li et al. proposed a semi-supervised uncertainty-guided dual-consistency learning segmentation network (UDC-Net) that imposes image transformation equivalence and feature perturbation invariance to effectively harness the knowledge from unlabeled data \citep{li2021dual}. The segmentation uncertainty was then quantified in two forms: confidence uncertainty calculated by the entropy of the mean prediction of multiple perturbated inputs, and consensus uncertainty quantified by the standard deviation over the multi-decoders’ predictions.

\paragraph{Uncertainty-aware learning} In 2020, Cao et al. presented an uncertainty-aware temporal ensembling model for semi-supervised breast mass segmentation \citep{cao2020uncertainty}. A temporal ensembling segmentation model was designed to segment breast mass using a few labeled and a large number of unlabeled images and an uncertainty map was estimated from MCD for each image; an adaptive ensembling momentum map and an uncertainty-aware unsupervised loss was designed and integrated with the temporal ensembling model.
\begin{table*}
  \centering
  \caption{Hybrids uncertainty quantification for medical image segmentation. The semi-supervised methods are highlighted in blue.}
  \scalebox{0.9}{
  \begin{tabular}{lllllll}
 \hline
Publications& Uncertainty  & Number of  &Clinical applications \\
&  methods& dataset  & \\
\hline

\cite{eaton2018towards} &MC sampling, TTD & 1&Brain tumour segmentation\\
\cite{dhamala2018quantifying}&MCMC, PDF&2&Cardiac electrophysiology segmentation\\
\cite{wang2019aleatoric} &MCD, TTA &1&Fetal brain and brain tumor segmentation\\
\cite{jungo2019assessing} &MCD, Ensemble&2&Brain tumor and skin lesion segmentation \\
\cite{jungo2020analyzing}&MCD, Ensemble &1 &Brain tumor segmentation\\
\textcolor{blue}{\cite{venturini2020uncertainty}}& TTA, TTD& 2 & Hippocampal and fetal brain segmentation\\
\cite{zheng2020cartilage}& Bootstrap, Ensemble&1&Cartilage segmentation \\
\cite{wang2020uncertainty}& MCD, Ensemble, BNN &1&Fetal brain segmentation\\
\cite{mehrtash2020confidence} &MCD, Ensemble&5& Brain tumor, ventricular and prostate segmentation\\
\cite{czolbe2021segmentation}&MCD, Ensemble, TTA& 2&Skin lesion\& lung cancer segmentation \\
\cite{mehta2021propagating}&MCD, Deep Ensemble, Ensemble-MCD &2& Lesion detection and brain tumour segmentation\\
\cite{zheng2021continual}&MC sampling, PD &3&Skin lesion segmentation \\
\cite{lin2022quality}&Fuzzy set, TTA&1&Skin lesion segmentation\\
\cite{lin2022novel} &MCD, fuzzy set, TTA&5&Skin lesion, nuclei, lung, breast, and cell segmentation\\
\cite{pandey2022can}&MCD, Ensemble, TTA &1 &Ultrasound bone segmentation\\
\cite{rajaraman2022uncertainty}&MCD, Interval analysis &1 &Tuberculosis segmentation\\
\cite{ng2022estimating}&MCD, Ensemble &2&Cardiac Segmentation\\
\cite{sagar2022uncertainty}&MCD, Ensemble, Ensemble-MCD&1&Brain tumor segmentation\\
\cite{ammari2023deep}&MCD, TTA, Shannon entropy&2&Right ventricular segmentation\\
\hline
\end{tabular}
}
\label{tab: hybrids}
\end{table*}
\subsubsection{Non-probabilistic methods}
Compared to the probabilistic-based method to quantify uncertainty due to lack of annotation, there are only a few non-probabilistic researches that study uncertainty in semi-supervised medical image segmentation frameworks. 

In 2022, Venturini et al. proposed an uncertainty-based method to improve the performance of segmentation networks when limited manual labels and estimated segmentation uncertainty on unlabeled data using TTA and TTD \citep{venturini2020uncertainty}. In the same year, Xiang et al. proposed a medical image segmentation framework that combines epistemic uncertainty-guided unsupervised learning and aleatory uncertainty-guided supervised learning with the ensemble of decoders \citep{xiang2022fussnet} Adiga et al. estimated the pixel-level uncertainty by leveraging the labeling representation of segmentation into a set of plausible masks \citep{adiga2022leveraging}.

In 2023, Huang et al. addressed the uncertainty caused by the low quality of the images and the lack of annotations using DST and deep learning \citep{huang2023semi} with a semi-supervised learning algorithm proposed based on an image transformation strategy, a probabilistic deep neural network and an evidential neural network used in parallel to provide two sources of segmentation evidence, and Dempster’s rule used to combine the two pieces of evidence and reach a final segmentation result.

In the same year, Xu et al. proposed a dual uncertainty-guided mixing consistency network with a contrastive training module that improves the quality of augmented images by retaining the invariance of data augmentation between original data and their augmentations \citep{xu2023dual}. The dual uncertainty strategy calculates dual uncertainty obtained from $N$ stochastic forward passes with random dropout between two models to select a more confident area for subsequent segmentation. The mixing volume consistency module guides the consistency between the volume before and after segmentation using dual uncertainty. 

\section{Discussion}
\label{sec: discuss}
In this section, we first list the key insights of applying uncertainty quantification in MIA and discuss the limitations. We then identify some potential future research points for readers' convenience. 

\subsection{Uncertainty quantification methods}
First, the large number of studies incorporating uncertainty quantification in their medical analysis pipeline proves that the need to quantify uncertainty is well taken into account by the AI research community, showing that efforts are being made to bridge the gap between scientific research and clinical applications. 

Bayesian inference, although providing a strong theoretical background for uncertainty, is scarcely implemented for medical image analysis because of the requirement for the modification of the NN weights and the training paradigm, as well as the slow convergence tends \citep{osawa2019practical} and noisy gradient descent \citep{jospin2022hands} in complex scenarios.

MC methods tended to be the most popular approach for uncertainty quantification in MIA, representing around half of the implemented methods. This popularity can be explained by its easy implementation in a large majority of neural networks trained with dropout. However, MC sampling requires multiple inferences for the same input image, considerably extending the inference time, which may not be compatible with high requirements in clinical efficiency.

Model ensemble is a popular trick to improve predictive performance while also providing quality uncertainty estimates. Similar to MC methods, it also has drawbacks in computational cost and efficiency. 

Though the above probabilistic methods have gained enough attention in MIA and have achieved promising performance in estimating Out-of-Distribution (OoD) uncertainty when the model faces inputs that fall outside the range or distribution of the training data,  their limitations still remain when addressing or representing complex scenarios, e.g., In-Distribution (ID) uncertainty that arises from the inherent variability and noise within the dataset. For example, in the case of a multiclass problem (a three-class classification task ($\Omega=\{a, b, c\}$) as an example here), a good uncertainty model should be able to model the possible intermediate classes between the totally certain and totally uncertain about a class (i.e., any subset of $\Omega$, e.g., $\{a, b\}$, $\{b, c\}$.), depending on the informativeness of the training data with respect to the class membership of the pattern under consideration \citep{denoeux2000neural}. Take three disease diagnoses as an example: an expert confirms that the patient does not have disease $a$ but may have disease $b$ or $c$; a good uncertainty model should then have the ability to model such ID uncertainty in an informative way, i.e, the degree of belief or plausibility that the patient be classified in to subset $\{b, c\}$. In practical scenarios, standard probabilistic uncertainty approaches, such as MCD or Ensemble, often fall short of effectively quantifying ID uncertainty. These approaches attempt to capture ID uncertainty by generating a set of predictions and calculating statistical indicators such as variance, offering only a singular uncertainty value without further context. Consequently, this limitation hampers the effectiveness of probabilistic methods in modeling ID uncertainty \citep{snoek2019control, ulmer2021know}.

Non-probabilistic methods attract people's attention in modeling fuzzy, noisy, or uncertain information and motivate the development of methods tailored for uncertain both ID and OoD. Compared with the probabilistic uncertainty methods, non-probabilistic uncertainty quantification methods release the requirement of strong assumptions about the real distribution and modeling uncertainty based on fuzzy or soft conception. DST, the most popular non-probabilistic uncertainty method, can model OoD uncertainty with full ignorance about prediction and model ID uncertainty by providing comprehensive belief and plausibility context about any subset of $\Omega$. Besides uncertainty quantification, DST also offers a way to combine multiple unreliable information, which is particularly useful in fusing multi-modality or cross-modality medical image data \citep{huang2022evidence}. Moreover, the introduction of DST with neural networks, i.e., EKNN \citep{denoeux1995k} and ENN \citep{denoeux2000neural}, makes it possible to integrate DST with SOTA deep learning models and, therefore, popularized its application in MIA. Other non-probabilistic uncertainty methods, such as fuzzy sets and fuzzy logic theory, interval analysis, and test time augmentation, although less frequently mentioned as DST, are also good choices for uncertainty quantification and can be further studied to integrate them with SOTA deep learning models. 

\subsection{Evaluation criteria} 
According to our literature review, a large variety of evaluation protocols are reported to assess the quality of uncertainty estimation. In the context of MIA, if multiple manual expert delineations are available for a given input image, the inter-rater variability is usually used as ground truth uncertainty to be compared with the predicted one. The related research has gained promising achievement and contributed to the development of uncertainty estimation in MIA. However, most of the time, the corresponding uncertainty values are not provided. Thus, evaluating uncertainty results relies on proxy tasks, such as detecting sample variance, predictive entropy, misclassification, OoD, or calibration performance. One possible evaluation method is to determine whether performing a task that takes uncertainty into account improves the performance calculated on the criteria dedicated to this task, for example, the Dice coefficient for a segmentation task. These methods are inspired by concrete applications of uncertainty in a real-world scenario. 

However, while several metrics exist to evaluate uncertainty estimation methods, none capture the complete picture. Metrics like calibration and coverage probability provide insights into specific aspects of uncertainty estimation but may not fully capture other important characteristics, such as the ability to capture epistemic and aleatory uncertainty separately. Therefore, we suggest researchers take task-depended clinical expectations/requirements into consideration when choosing uncertainty quantification evaluation criteria and ensure the fairness and pertinence of the evaluation criteria. 

\subsection{Applications}
Analyzing uncertain information in image reconstruction and registration can improve the quality of medical images. Uncertainty quantification assesses the impact of radiation dose or contrast agent usage on reconstructed images and can help find the most optimizing imaging condition. Medical image registration involves aligning and transforming multiple images to enable comparison or fusion. Uncertainty estimates help understand the confidence level of the registration process. This is important when the alignment is challenging due to image noise, artifacts, or deformations, especially for multi-modal medical image registration tasks. 

In medical diagnosis, using a detection, classification, or segmentation model developed from an imbalanced dataset (which is a common situation in the medical domain) is risky because the model might be overconfident or overconfident. Uncertainty estimation can thus be used to identify where pixel/voxel or object-level predictions are less certain, therefore helping clinicians understand the reliability of the prediction results and identify areas where automatic prediction may fail and manual intervention might be necessary by providing insights into regions of high ambiguity or uncertainty. This can be particularly useful in minimizing false positives and false negatives and detecting Out-of-Distribution or ambiguous In-Distribution samples that might need specialized handling. Apart from disease diagnosis, prediction of treatment outcomes or disease development is also important to improve the cure rate. Uncertainty estimates provide insights into the range of possible outcomes, supporting personalized treatment strategies and allowing researchers to set realistic expectations for model performance.

To conclude, uncertainty quantification provides critical information about the reliability and confidence of the analysis. This information is particularly valuable in medical applications due to the critical nature of the decisions made based on these predictions, impacting patient care and treatment outcomes. By incorporating uncertainty estimation, MIA becomes more transparent, trustworthy, and aligned with the clinical workflow, which helps bridge the gap between artificial intelligence algorithms and clinical practice, enhancing the acceptance and trustworthiness of AI-assisted medical decisions. Furthermore, building public trust will also help to improve the general fairness of AI healthcare systems.

Apart from the methods mentioned above that focus on studying the uncertainty of the medical image analysis results, a branch of literature also focuses on modeling or analyzing the uncertainty of image labels itself. Medical experts may have varied interpretations of the same image, leading to inter-observer variability \citep{vinod2016review, jungo2018effect}. Additionally, the same expert may interpret an image differently on different occasions, causing intra-observer variability \citep{sampat2006measuring, schmidt2023probabilistic}. Such discrepancies in annotations introduce uncertainty and complexity in medical image analysis. Therefore, the label uncertainty modeling approaches focus on such datasets, and studying effective methods for modeling and reducing the inter-observer and intra-observer variability is necessary and important. There are some researches that take into account medical image labeling uncertainty, which can be classified according to the focus on inner uncertainty or inter-observer uncertainty modeling, i.e., image label uncertainty modeling and fusion of uncertain image labels. Moreover, there are some researchers who contribute to open-source new datasets with uncertain ground truth. Readers can refer to Supplementary Material B for related analysis. 

\subsection{Perspectives}
Based on the discussion of the advantages and limitations of existing uncertainty quantification methods, we suggest several future research points to further improve the implications of uncertainty quantification in MIA. 

\paragraph{Effectiveness}
The most critical limitation of present uncertainty quantification research is the lack of ground truth uncertainty, leading to the lack of standardized evaluation metrics for uncertainty quantification methods. The uncertainty associated with ground truth labels can propagate and affect model uncertainty estimates. However, ground truth labels are not always definitive due to inherent inter-observer variability, ambiguous cases, or inherent limitations of manual annotations. Moreover, the lack of the uncertainty ground truth limits the understanding of sources and reasons behind uncertainty and the explanation of uncertainty to clinicians or users. Though some researchers use inter-rater variability as uncertainty ground truth, it is still unclear whether it is theoretically guaranteed. For example, for a segmentation task, experts can somewhat give random variations around the boundaries of the target object, over-segment, or alternatively under-segment the same object of interest based on their annotation style. This inter-rater variability is thus instead linked to contextual biases (e.g., radiologist experience or annotation habits) rather than to the true uncertainty of the label \citep{mehta2022qu}. Therefore, we encourage researchers to put efforts into constructing MIA datasets with both accuracy and uncertain ground truth and set up standardized evaluation metrics for uncertainty quantification methods. A simple solution can be providing diagnosis/detection/prediction/segmentation/classification ground truths as well as providing a corresponding confidence index. 
 
\paragraph{Explainability}
SOTA uncertainty quantification methods, such as deep learning ensembles or MCD, may lack interpretability, making it challenging to explain the uncertainty estimation process to clinicians or patients. Therefore, the link between explainability and uncertainty would be interesting to study. Studying the relationship allows us to understand both how the prediction is made and whether or not it should be trusted, in other words, whether or not the results are reliable. An interesting research point would be to complement uncertainty estimates with explanations, helping the user understand the uncertainty of each source and how the uncertain sources are summarized and summed to reach a final decision. For example, in \citep{HUANG2023737}, Huang et al. proposed a deep evidential fusion framework with uncertainty quantification and contextual discounting for multimodal medical image segmentation. This approach is the first attempt to explain the decision-making process by quantifying subject-level uncertainty with contextual discounting to the fusion of deep neural networks and applying it to multimodal medical image segmentation tasks. Another potential research work is studying the relationship between uncertainty and reliability. Conventional research typically treats uncertainty as an opposite indicator of reliability, \citep{modarres2016reliability, ovadia2019can}, i.e., the lower the uncertainty, the higher the reliability, which is just an approximation and has limitations in explaining more complex situations such as uncertain but reliable models. Therefore, integrating uncertainty with reliability, i.e., studying the relationship between uncertainty and reliability, could also be an exciting and significant subject.

\paragraph{Efficiency}
As shown, the vast majority of the implemented uncertainty quantification methods are based on a sampling protocol, such as MCD and Bayesian inference, aiming at generating multiple predictions. However, they can be computationally expensive and time-consuming, which, therefore, limits their practical application in real-time or clinical settings, where quick and efficient analysis is crucial. The recently popular deep ensemble models, their superior uncertainty measure, along with the high computational cost. Non-probabilistic methods, such as DST, compute the uncertainty in a quick and efficient manner that requires only a single forward step, which is generally required for medical applications, indicating a promising direction to be further explored. 

\paragraph{Clinical applications}
Integrating uncertainty quantification into clinical workflows and decision-making processes can be challenging due to the limited trust in existing ML models and the limited clinical validation. Therefore, careful consideration and adaptation of uncertainty quantification are required to align research with clinical guidelines and to fit it within the clinical context. We thus suggest researchers integrate clinical validation and take ethical and legal problems into consideration when developing their MIA models to 1) enable more reliable, interpretable, and applicable uncertainty quantification models; 2) ensure their clinical utility, interpretability, and impact on patient outcomes; 3) ensure their fairness to the public.
 
\section{Conclusion}
\label{sec: conclu}
This review provides an overview of the uncertainty quantification methods commonly implemented in machine learning-based medical image applications. Numerous phenomena can cause predictive uncertainty, such as noisy images, imperfect ground truth labels, incomplete data, and inter-site image variability. The literature proposes various methods to quantify uncertainty applied to an extensive range of medical image applications. As demonstrated in this review, developing trustable AI solutions integrating uncertainty quantification of the computed predictions is an active search topic that has many potential future directions.

\section*{Acknowledgments}
This research is supported by A*STAR, CISCO Systems (USA) Pte. Ltd, and National University of Singapore under its Cisco-NUS Accelerated Digital Economy Corporate Laboratory (Award I21001E0002) and the NMRC Health Service Research Grant (MOH-000030-00).

\bibliographystyle{model2-names.bst}\biboptions{authoryear}
\bibliography{refs}

\section*{Supplementary Material A}

\subsection*{Bayesian inference} 
\paragraph{Probabilistic Distribution (PD)}
In Bayesian inference, probabilistic distribution, such as Gaussian distribution (the most commonly used one), Beta distribution, Poisson Distribution, Exponential distribution, and Dirichlet distribution, are usually used to generate distributions over predictions rather than point estimates \citep{wallman2014computational, liao2019modelling, islam2021spatially}. The parameters of the posterior probabilistic distribution provide estimates of the parameter of interest, and the posterior covariance matrix gives the parameters' uncertainties. The diagonal elements of the covariance matrix correspond to the variances of the estimated parameters.

\paragraph{Gaussian Process (GP)}
GP is a non-parametric approach used to model functions as probability distributions over possible functions \citep{wachinger2014gaussian, wu2021quantifying, peter2021uncertainty}. GP provides not only point predictions but also the associated uncertainty estimates at every point in the input space, making them valuable for regression, interpolation, and optimization tasks where uncertainty needs to be considered.

\paragraph{Bayesian Neural Networks (BNNs)}
With the success of neural networks (NNs), Bayesian inference is also integrated into neural networks to contract a BNN for uncertainty estimation \citep{blundell2015weight, bian2020uncertainty, li2021uncertainty, krygier2021quantifying}. In BNN, each weight $w$ of the NN is replaced by placing a prior distribution over the neural network weights rather than having a single fixed value. A prior distribution $p(w)$ is first initialized over the NN weights and the model learns the posterior distribution $p(w|D)$ given the training dataset $D$ and the prior distribution during training. The trained BNN is akin to a virtually infinite ensemble of NNs, where each instance has weights drawn from the learned posterior distribution. 

\subsection*{MC methods} 
\paragraph{MC sampling}
MC sampling \citep{zheng2021continual, ghoshal2021cost}is a general interpretation of methods that estimates uncertainty by drawing random samples from a given distribution (normally Gaussian distribution), estimating quantities of interest, and characterizing uncertainty using the obtained samples. Two basic sampling types: simple sampling which draws independent samples from the distribution of interest and importance sampling which draws samples from a different, easier-to-sample distribution and uses weights to adjust for the difference between the true distribution and the sampling distribution are used. Advanced techniques such as Latin hypercube sampling and Jackknife resampling, are also employed to enhance the efficiency of MC methods and reduce the number of required samples.

\begin{figure*}
    \centering
    \includegraphics[width=0.8\textwidth]{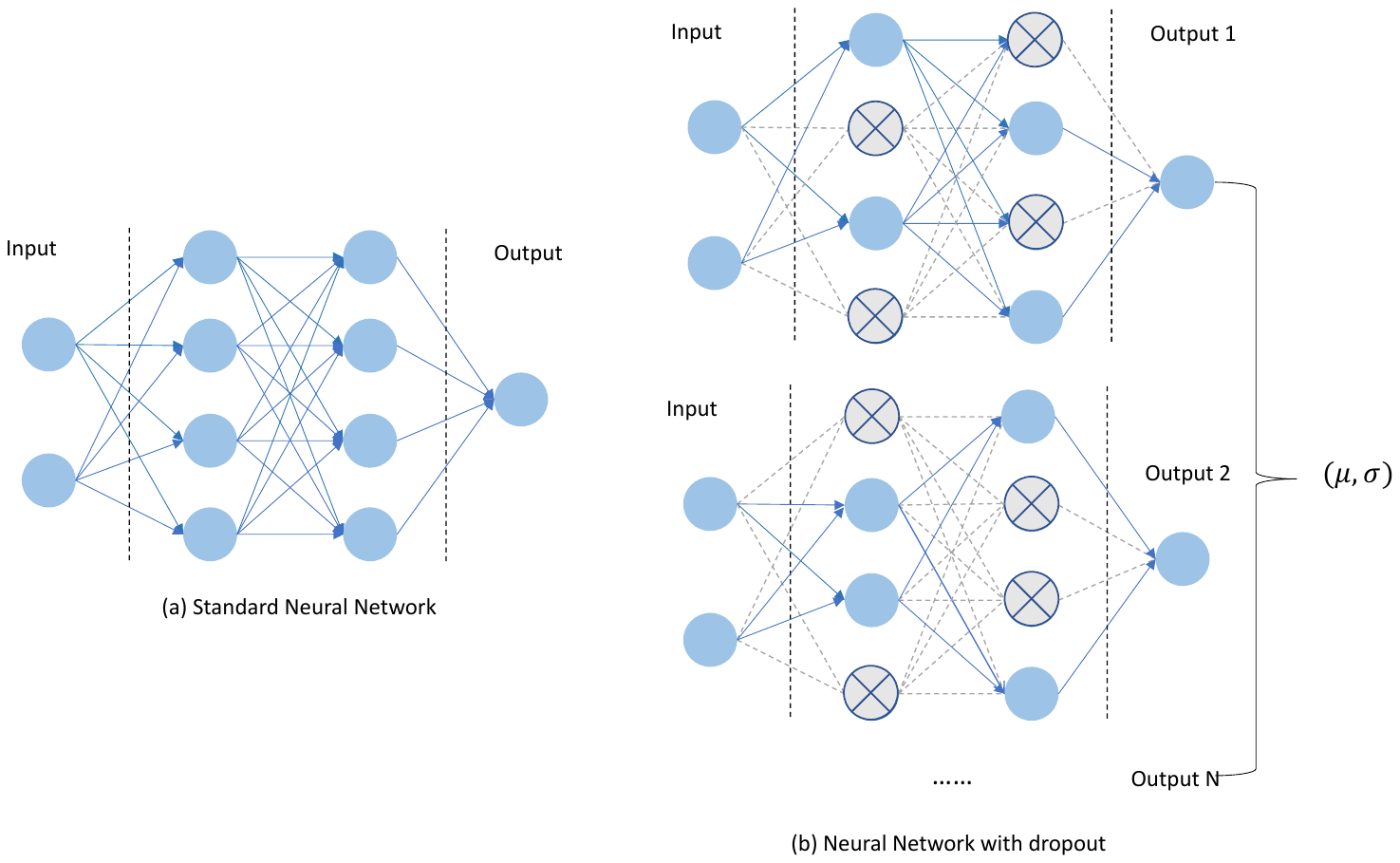}
    \caption{Example of (a) standard Neural Network and (b) Neural Network with Dropout. The dropped neurons were marked in grey and linked by a dotted line. Parameter $\mu$ is the mean or expectation of $N$ distributions, and $\sigma$ is its standard deviation.}
    \label{fig: dropout}
\end{figure*}

\paragraph{Test-Time Dropout (TTD)}
Dropout \citep{srivastava2014dropout} is primarily a regularization technique used during training to prevent overfitting in neural networks. However, it can also be adapted for uncertainty quantification during the test or inference phase. Test-time dropout (TTD) is commonly used in various machine learning applications to estimate predictive uncertainty and make probabilistic predictions. Figure \ref{fig: dropout} shows an example of a standard Neural Network (left) and a Neural Network with Dropout (right), where the dropped neurons were marked in grey and linked by a dotted line. By applying TTD, the model generates different predictions for the same input data, and these predictions reflect the uncertainty associated with the model's weights and architecture.

\paragraph{Monte Carlo dropout (MCD)}  
In \cite{gal2016dropout}, the authors demonstrated that an NN trained with dropout operation \ref{fig: dropout}(b) is able to efficiently approximate Bayesian inference that sampling from a variational family (Gaussian Mixture) and approximate the true deep Gaussian process posterior without the associated prohibitive computational cost. Based on this principle, MCD, a SOTA technique for estimating uncertainty in predictions, is proposed. In MCD, dropout is applied at both training and test time. During test time, multiple forward passes are performed with dropout instead of using a single forward pass, resulting in a collection of different predictions for each input. 

\begin{figure}
    \centering
    \includegraphics[scale=0.3]{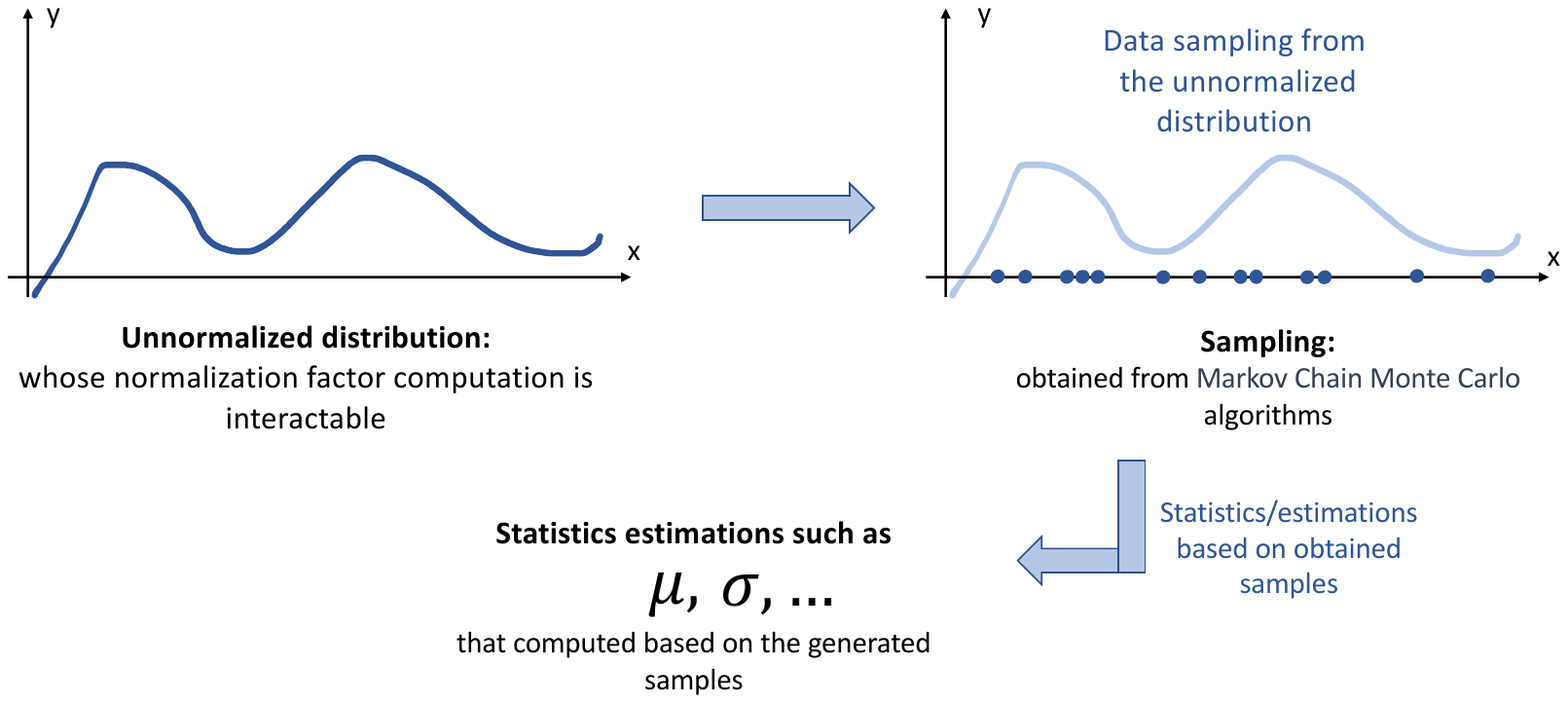}
    \caption{Markov Chain Monte Carlo reasoning process. Parameter $\mu$ is the mean or expectation of all sampling samples, $\sigma$ is the corresponding standard deviation.}
    \label{fig: mcmc}
\end{figure}

\paragraph{Markov Chain Monte Carlo (MCMC)} 
MCMC methods use Markov chains to generate dependent data samples. The basic idea is to build such Markov chains, which are easy to sample from and whose stationary distribution is the target distribution, such that when following them, in the limit, we obtain samples from the target distribution \citep{Christophe2023}. MCMC methods, such as the Metropolis-Hastings algorithm \citep{chib1995understanding}, Gibbs \citep{kozumi2011gibbs} or slice sampling \citep{neal2003slice}, are used to sample from probability distributions. These methods are particularly useful when analytical solutions are not available. Figure \ref{fig: mcmc} shows the MCMC reasoning process. 

\paragraph{Bootstrap}
Bootstrap \citep{efron1992bootstrap, davison1997bootstrap}, a statistical technique used for uncertainty quantification by estimating the sampling variability of a statistical estimator or model, also belongs to the broader category of MC sampling. It involves resampling the observed data (with replacement) to create multiple bootstrap samples. Those samples are then used to estimate the uncertainty by calculating statistics such as the standard deviation, confidence intervals, or percentile intervals of interest. 
\begin{itemize}
    \item Step 1 (Sampling): Randomly select a bootstrap sample of size $N$ (with replacement) from the original dataset.
    \item Step 2 (Estimation): Apply the desired estimation or modeling procedure to the bootstrap sample to obtain an estimate of interest. 
    \item Step 3: Repeat Steps 1 and 2 $N$ times (typically, $N \gg D$ ), each time generating a new bootstrap sample and computing the corresponding estimation.
    \item  Step 4 (Uncertainty calculation): Analyze the distribution or variability of the obtained estimates across the $N$ bootstrap samples. 
\end{itemize}

\subsection*{Deep ensemble} 

\begin{figure*}
    \centering
    \includegraphics[width=0.5\textwidth, angle=90]{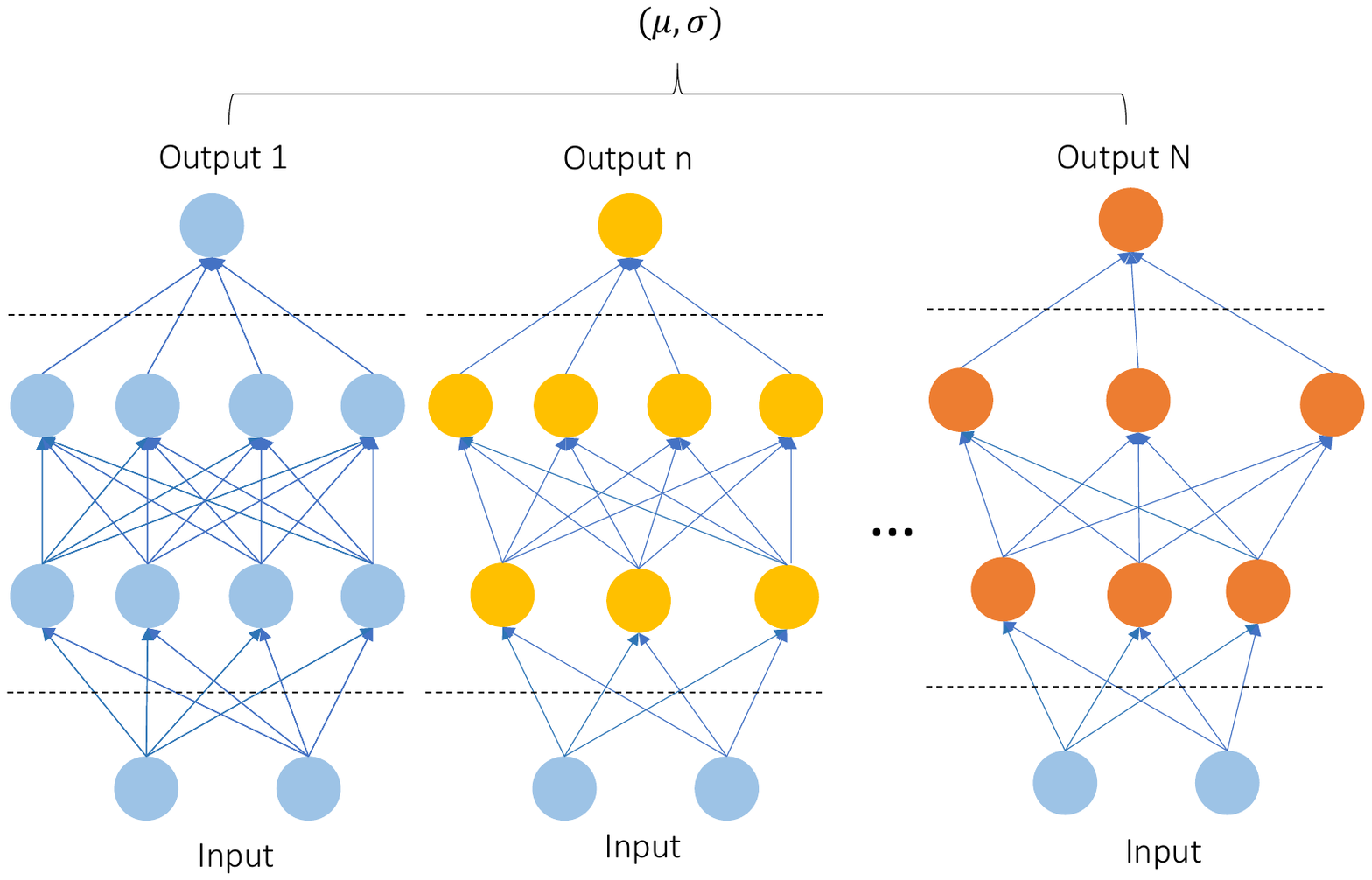}
    \caption{Example of an ensemble model with multiple neural networks. Parameter 
$\mu$ is the mean or expectation of $N$ distributions, $\sigma$  is the standard deviation. }
    \label{fig: ensemble}
\end{figure*}

The idea of deep ensemble is that $N$ neural networks are trained independently to collect $N$ deterministic predictions. The variability in predictions across ensemble members is then used to estimate uncertainty \citep{ guo2022cardiac, zhang2023multi}. Figure \ref{fig: ensemble} shows an example of an ensemble model with multiple neural networks, where epistemic uncertainty is captured as different models in the ensemble may have different learned representations, reflecting uncertainty about the true model structure.

\subsection*{Dempster-Shafer Theory} 
Let $\Omega =\{\omega _{1},\omega _{2}, ..., \omega_{C}\} $ be a finite set of all possible hypotheses about some problem, called a frame of discernment. Evidence about a variable $\omega$ taking values in $\Omega$ can be represented by mass function $m$, from the power set $2^{\Omega}$ to $[0, 1]$, such that
\begin{subequations}
\begin{align}
    \sum _{A\subseteq \Omega }m(A)=1,\\
    m(\emptyset)=0.
    \label{eq:evidence}
\end{align}    
\end{subequations}
Each subset $A \subseteq \Omega$ and $m(A)$ is called a focal set of $m$. The uncertainty (total ignorance) of the problem can be represented as $m(\Omega)$. In DST, the belief about a certain item is elaborated by aggregating different belief functions over the same frame of discernment. 

\paragraph{Shafer's model} 
Assuming that conditional density functions $f(x\mid \omega_c)$ are known, then the conditional likelihood associated with the pattern $X$ is defined by $\ell (\omega_c\mid x)=f(x\mid \omega_c)$. The mass functions are defined according to the knowledge of all hypotheses $\omega_1, \ldots, \omega_C $. Firstly, the plausibility of a simple hypothesis $\omega_c$ is proportional to its likelihood. The plausibility is, thus, given by
\begin{equation}
    Pl(\{\omega_c\})=\hslash \cdot \ell (\omega_c\mid x), \quad  \forall \omega_c \in \Omega,
    \label{eq:14}
\end{equation}
where $\hslash $ is a normalization factor with $\hslash=1/  \max_{\omega\in\Omega} \ell(\omega \vert x)$. The plausibility of a set $A$ is, thus, given by
\begin{equation}
  Pl(A)=\hslash\cdot \underset{\omega_c \in A}{\max} \ell (\omega_c\mid x).
\end{equation}

\paragraph{Dempster's rule} Given two mass functions $m_{1}$ and $m_{2}$ derived from two independent items of evidence, the final belief that supports $A$ can be obtained by combining $m_{1}$ and $m_{2}$ with Dempster's rule \citep{shafer1976mathematical}, which is defined as 
\begin{equation}
    (m_{1}\oplus m_{2})(A)=\frac{1}{1-\kappa }\sum _{B\cap D=A}m_{1}(B)m_{2}(D),
    \label{eq:demp1}
\end{equation}
for all $A\subseteq \Omega, A\neq \emptyset$, and $(m_{1}\oplus m_{2})(\emptyset)=0$. The coefficient $\kappa$ is the degree of conflict between $m_{1}$ and $m_{2}$ that
\begin{equation}
    \kappa=\sum _{B\cap D=\emptyset}m_{1}(B)m_{2}(D).
    \label{eq:demp2}
\end{equation}

\paragraph{Evidential K-Nearest Neighbor (EKNN) rule}
Let $N_K(x)$ denote the set of the $K$ nearest neighbors of $x$ in a learning set. Each $x_{i}\in N_K(x)$ is considered as a piece of evidence regarding the class label of $x$. The strength of evidence decreases with the distance between
$x$ and $x_{i}$. The evidence of $(x_{i},y_{i})$ support class $c$ is represented by
\begin{subequations}
\begin{align}
m_i(\{\omega_{c}\})&=\varphi _{c}(d_{i})y_{ic}, \quad  1 \le c \le C,\\
m_{i}(\Omega)&=1-\varphi_{c}(d_{i}),
\end{align}
\label{eq:18}
\end{subequations}
where $d_{i}$ is the distance between $x$ and $x_{i}$, which can be the Euclidean or other distance function; and $y_{ic}=1$ if $y_{i}=\omega_{c}$ and $y_{ic}=0$ otherwise. Function $\varphi _{c}$ is defined as
\begin{equation}
    \varphi _{c}(d)=\alpha \exp(-\gamma d^{2}),
    \label{eq:19}
\end{equation}
where $\alpha$ and $\gamma$ are two tuning parameters. The evidence of the $K$ nearest neighbors of $x$ is fused by Dempster's rule:
\begin{equation}
m=\bigoplus _{x_{i}\in N_K(x)}m_{i}.  
\label{eq:20}
\end{equation}
The final decision is made according to maximum plausibility. 

\paragraph{Evidential C-Means (ECM)}
In \citep{denoeux2004evclus}, Denoeux et al. proposed an evidential clustering algorithm that extends the notion of fuzzy partition with \emph{Credal partition}, which extends the existing concepts of hard, fuzzy (probabilistic), and possibilistic partition by allocating each object a 'mass of belief,' not only to single clusters but also to any subsets of $\Omega= \{\omega_{1}, ..., \omega_{C}\}$. Based on the credal partition, Evidential C-Means (ECM) \citep{masson2008ecm} was introduced to generate mass functions. In ECM, a cluster is represented by a prototype $p_{c}$. For each non-empty set $A_{j}\subseteq\Omega$, a prototype $\bar{p_{j}}$ is defined as the center of mass of the prototypes $p_c$ such that $\omega_c\in A_j$. Then the non-empty focal set is defined as $F=\{A_{1}, ..., A_{f}\}\subseteq 2^{\Omega}\setminus\left\{\emptyset\right \} $. Deriving a credal partition from object data implies determining, for each object $x_i$, the quantities $m_{ij}=m_i(A_j), A_i\ne \emptyset, A_j \subseteq \Omega$. The distance between an object and any nonempty subset of $\Omega$ has thus to be defined by
 \begin{equation}
    d_{ij}^2=\left \| x_{i}-\bar{p_{j}}\right \|^2.
    \label{eq:27}
 \end{equation} 

\paragraph{Evidential Neural Network (ENN)}
In \citep{denoeux2000neural}, Den{\oe}ux proposed an Evidential Neural Network (ENN) classifier in which mass functions are computed based on distances to prototypes. The ENN classifier is composed of an input layer of $H$ neurons, two hidden layers, and an output layer. The first input layer is composed of $I$ units, whose weights vectors are prototypes $p_1,\ldots, p_I$ in input space. The activation of unit $i$ in the prototype layer is
\begin{equation}
    s_i=\alpha _i \exp(-\gamma_i d_i^2),   
    \label{eq:si}
\end{equation}
where $d_i=\left \| x-p_i \right \| $ is the Euclidean distance between input vector $x$ and prototype $p_i$, $\gamma_i>0$ is a scale parameter,  and $\alpha_i \in [0,1]$ is an additional parameter. The second hidden layer computes mass functions $m_i$ representing the evidence of each prototype $p_i$, using the following equations: 
\begin{subequations}
\begin{align}
m_i(\{\omega _{c}\})&=u_{ic}s_i, \quad c=1,..., C\\
m_{i}(\Omega)&=1-s_i, 
\label{eq:9}
\end{align}
\end{subequations}
where $u_{ic}$ is the membership degree of prototype $i$ to class $\omega_c$, and $\sum _{c=1}^C u_{ic}=1$. Finally, using Dempster's rule, the third layer combines the $I$ mass functions $m_1,\ldots,m_I$. The output mass function $m=\bigoplus_{i=1}^{I} m_i$ is a discounted Bayesian mass function that summarizes the evidence of the $I$ prototypes. 

\paragraph{Subjective Logic (SL)}
Subjective logic \citep{josang2006trust, josang2016subjective} extends DST by introducing additional concepts and principles to handle subjective judgments and uncertainty. It incorporates degrees of belief, disbelief, and uncertainty to capture subjective opinions and incomplete information. Arguments in SL are subjective opinions about state variables that can take values from a domain (aka state space), where a state value can be thought of as a proposition that can be true or false. A binomial opinion applies to a binary state variable and can be represented as a Beta PDF (Probability Density Function) \citep{kotz2004continuous}. A multinomial opinion applies to a state variable of multiple possible values and can be represented as a Dirichlet PDF (Probability Density Function) \citep{olkin1964multivariate}. For each input $X_n$, the SL provides belief mass $b_c$ for different classes (Assuming $C$ classes here) and an uncertainty mass $U$ for whole classes. Accordingly,
\begin{equation}
 \sum_{c=1}^{C} b_c + u=1,
\end{equation}
where $b_c \ge 1$ and $u \ge 1$ denote the probability of the input $X_n$ for the $c^{th}$ class and the input's global ignorance (uncertainty). The evidence $e^n =\left [e^n_1,..., e^n_C\right ] $ for the classification result is acquired by an activation function layer softplus and $e^n_c \ge 0$. Then the Dirichlet distribution can be parameterized by $\alpha^n = \left [ \alpha^n_1, ..., \alpha^n_C \right ]$, which associated with the evidence $e^n_c$, i.e., $\alpha^n_c=e^n_c+1$. In the end, the image-level belief mass and the uncertainty mass of the classification can be calculated by:
\begin{equation}
b_c^n=\frac{e_C^n}{S^n} =\frac{\alpha_c^n-1}{S^n},
\end{equation}
and 
\begin{equation}
U^n=\frac{C}{S^n},
\end{equation}
where $S^n= {\textstyle \sum_{c=1}^{C}\alpha_c^n}= {\textstyle \sum_{c=1}^{C}e_c^n+1}$ represents the Dirichlet strength.

\section*{Supplementary Material B}

Table \ref{tab: label uncertianty} lists the related works that focus on medical image labeling uncertainty analysis. 
\begin{table*}
  \centering
  \caption{Label uncertainty modeling\& analysis in medical image analysis}
  \scalebox{0.85}{
  \begin{tabular}{llllllll}
 \hline
&Publications & Methods to uncertain   & New & Clinical applications  \\
& & label analysis & dataset &  \\
\hline
&\cite{kohl2018probabilistic} &Plausible sets &No&Lung abnormalities segmentation \\
&\cite{liao2019modelling}& PD & No& 2D echo quality assessment\\
& \cite{czolbe2021segmentation}& Ensemble, MCD, TTA & No& Skin lesion\& lung cancer segmentation \\
&\cite{pham2021interpreting}& Soft label & No&Thoracic diseases classification\\
&\cite{redekop2021uncertainty}&TTD&No& Skin lesion and liver segmentation\\

 &\multirow{2}{*}{\cite{islam2021spatially}} & \multirow{2}{*}{PD} &\multirow{2}{*}{No}&Brain tumors, prostate zones, kidney tumors \\
&&&&and lung nodules segmentation\\
&\cite{peter2021uncertainty}&PD & No &chest CT scan registration\\
\multirow{1}{*}{Label} &\cite{khawaled2022npbdreg}& PD &No& Brain MRI registration \\
\multirow{1}{*}{uncertainty} &\cite{adiga2022leveraging}& PD &No& Left atrium segmentation\\

\multirow{1}{*}{modeling} &\cite{wu2021uncertainty}&MCD& No& Mitochondria segmentation\\
&\cite{ghoshal2022calibrated}&MCD&No& COVID-19 detection\\
&\cite{aljuhani2022uncertainty}&MCD&No&Tumor region classification\\ 
&\cite{javadi2022towards}&TTA, TTD&No&Prostate cancer detection\\

&\cite{wu2023upl}&TTA&No&Fetal brain Segmentation\\
&\cite{islam2023paced}& PD &No&Breast segmentation\\
&\cite{del2023labeling}&Soft label&No&  Histology image classification\\

\hline
\multirow{1}{*}{Uncertain}& \cite{jungo2018effect}&STAPLE,  vote, intersection, union&  No& Brain tumor segmentation\\
label&\cite{li2022ultra}& multi-rater label fusion 

&No& Breast tumor cellularity assessment estimation\\
fusion &\cite{lemay2022label}&STAPLE, average, sampling &No& Spinal cord gray matter, brain lesion segmentation\\
\hline
\multirow{2}{*}{New dataset}& \cite{irvin2019chexpert}&  PD   &Yes&  Chest radiograph interpretation\\
&\multirow{2}{*}{\cite{ju2022improving}}&\multirow{2}{*}{MCD}&\multirow{2}{*}{Yes} & Skin lesions, prostate cancer \\
& &&&and retinal disease classification\\
\hline
\end{tabular}
}
\label{tab: label uncertianty}
\end{table*}
\subsection*{Image label uncertainty modeling}
\label{subsubsec: uncertainy-modeling}
To deal with the uncertainty of image labels, the straightforward way is to model it with a label distribution map using fuzzy concepts. It can be achieved by introducing probabilistic uncertainty modeling algorithms such as prediction variability \citep{liao2019modelling} or non-probabilistic algorithms such as fuzzy predictions \citep{kohl2018probabilistic, adiga2022leveraging} and label smoothing strategies \citep{del2023labeling, islam2021spatially, pham2021interpreting}.  

In 2018, Kohl et al. approximated the uncertain expert label distribution using generative neural networks in MIA task \citep{kohl2018probabilistic}. They proposed a generative segmentation model based on a combination of a U-Net with a conditional variational autoencoder that is capable of efficiently producing an unlimited number of plausible sets. 

In 2019, Liao et al. proposed a method to model the intra-observer variability in echo quality assessment as an aleatoric uncertainty modeling regression problem with Cumulative Density Function (CDF) Probability \citep{liao2019modelling}. It addressed the observer variability as aleatoric uncertainty, which models experts’ opinions as Laplace or Gaussian distributions over the regression space. 

In 2021, Czolbe et al. considered four established strategies, i.e., U-Net with Softmax Output, Ensemble Methods, MCD and Probabilistic U-Net to address the inter-observer variability or intra-observer variability \citep{czolbe2021segmentation}. In the same year, Pham et al. presented a multi-label classification framework based on deep CNNs for predicting the presence of 14 common thoracic diseases and observations \citep{pham2021interpreting}. They trained several state-of-the-art CNNs that exploit hierarchical dependencies among abnormality labels using the label smoothing technique to handle uncertain samples. Redekop and Chernyavskiy proposed to train binary segmentation DCNNs using sets of unreliable pixel-level annotations \citep{redekop2021uncertainty}. Islam et al. proposed a spatially varying label smoothing mechanism for incorporating structural label uncertainty by capturing ambiguity about object boundaries in expert segmentation maps in \citep{islam2021spatially}.

In 2022, Adiga et al. proposed to estimate the pixel-level uncertainty by leveraging the labeling representation into a set of plausible masks and estimating the uncertainty with a single inference from the labeling representation \citep{adiga2022leveraging}. In the same year, Aljuhani et al. presented an importance-based sampling framework with MCD-based approximate inference for robust histopathology image analysis \citep{aljuhani2022uncertainty}. Ghoshal et al. extended the approximate inference for the loss-calibrated Bayesian framework to drop weights-based Bayesian neural networks by maximizing expected utility over a model posterior to calibration uncertainty in deep learning \citep{ghoshal2022calibrated}. 

In 2023, Del Amor et al. designed an uncertainty-driven labeling strategy to generate soft labels from 10 non-expert annotators for multi-class skin cancer classification \citep{del2023labeling}. Based on the soft annotations, they proposed an uncertainty estimation framework to handle these noisy labels with a novel formulation using a dual-branch min–max entropy calibration to penalize inexact labels during the training. 
\subsection*{Fusion of uncertain image labels}
\label{subsubsec: uncertainy-fusion}

Research on the fusion of uncertain image labels mainly focuses on modeling and addressing the conflicts or ambiguities among labels. This part of the study deals only with the post-processing of uncertain labels, therefore we do not distinguish between probabilistic and non-probabilistic methods. In 2018, Jun et al. analyzed the effect of common image label fusion techniques with uncertain labels: (a) no fusion, (b) majority vote, (c) STAPLE \citep{warfield2002validation}, (d) intersection and (e) union of all observers, and then analysis model’s capability to learn the inter-observer variability into the estimation of segmentation uncertainty regardless of the image content in \citep{jungo2018effect}. An interesting finding is that the obtained results highlighted the negative effect of fusion methods applied in deep learning to obtain reliable estimates of segmentation uncertainty and showed that the learned observers’ uncertainty can be combined with current MCD-based models to characterize the uncertainty of the model’s parameters. 

In 2022, Lemay et al. compared three label fusion methods: STAPLE, average of the rater's prediction, and random sampling of each rater's prediction \citep{lemay2022label}. The results indicated conventional models trained with a Dice loss, with binary inputs and sigmoid/softmax final activate, were overconfident and underestimated the uncertainty associated with inter-rater variability. Conversely, fusing labels by averaging with the soft prediction framework led to underconfident outputs and overestimation of the rater disagreement. 

To efficiently leverage the label ambiguities, in 2022, Li et al. proposed an uncertainty-aware label distribution learning framework \citep{li2022ultra} by converting single-value labels to discrete label distributions and modeling the ambiguity among all possible labels. The framework then learned label distributions by minimizing the KL divergence between the predicted and ground-truth label distributions and mimicked the multi-rater fusion process in clinical practice with a multi-branch feature fusion module to further explore the uncertainties of labels. 

\subsection*{New image dataset with uncertainty annotation}
\label{subsubsec: new-dataset}
In addition to modeling or analyzing the label uncertainty in the existing open public dataset, there are some researchers who contribute to larger-scale medical datasets with uncertainty annotation. For example, in 2019, Irvin et al. presented a large dataset of chest radiographs called CheXpert, which features uncertainty labels and radiologist-labeled reference standard evaluation sets. This dataset consists of 224,316 chest radiographs of 65,240 patients labeled for the presence of 14 common chest radiographic observations \citep{irvin2019chexpert}. To our knowledge, this is the first dataset that provided both accuracy and uncertainty annotations. It helps the development and validation of chest radiograph interpretation models towards improving healthcare access and delivery worldwide. In 2022, Ju et al. released a large re-engineered database that consists of annotations from more than ten ophthalmologists with an unbiased golden standard dataset for evaluation and benchmarking \citep{ju2022improving}.

Those label uncertainty analysis methods could have a high impact in real-world applications, such as being used as clinical decision-making algorithms accounting for multiple plausible semantic segmentation hypotheses to provide possible diagnoses and recommend further actions to resolve the present ambiguities.

\end{document}